\documentclass[letterpaper,11pt]{article}
\usepackage{amssymb}
\usepackage{amsmath}
\usepackage{amstext}
\usepackage{graphicx,epsfig}
\usepackage{epsfig}
\usepackage{verbatim} 
\usepackage{color}
\usepackage{ulem}
\usepackage{enumitem}
\usepackage{subfigure}
\usepackage{bbm}
\usepackage{parskip}
\usepackage{dsfont}
\usepackage[numbers,sort&compress]{natbib}
\usepackage[body={6.5in, 9in}, right=1in, top=1in]{geometry}
\usepackage{mathtools}
\usepackage{comment}
\usepackage{caption}
\usepackage{bbold}
\usepackage{float}

\newcommand{\Comment}[1]{{}}
\definecolor{darkblue}{rgb}{0.15,0.35,0.55}
\definecolor{reddish}{rgb}{0.65, 0.2, 0.2}

\newcommand{\be}{\begin{equation}}
\newcommand{\ee}{\end{equation}}
\newcommand{\bea}{\begin{eqnarray}}
\newcommand{\eea}{\end{eqnarray}}
\newcommand{\beas}{\begin{eqnarray*}}
\newcommand{\eeas}{\end{eqnarray*}}

\def\({\left(}
\def\){\right)}

\def\gsim{ \lower .75ex \hbox{$\sim$} \llap{\raise .27ex \hbox{$>$}} }
\def\lsim{ \lower .75ex \hbox{$\sim$} \llap{\raise .27ex \hbox{$<$}} }

\usepackage{xcolor,colortbl}

\begin{document}
\def\thefootnote{\fnsymbol{footnote}}

\begin{center}
\LARGE{\textbf{The Equation of State of Dark Matter Superfluids}} \\[0.5cm]
 
\large{Anushrut Sharma, Justin Khoury and Tom Lubensky}
\\[0.5cm]

\small{
\textit{Department of Physics and Astronomy, University of Pennsylvania,\\ 209 South 33rd St, Philadelphia, PA 19104}}

\vspace{.2cm}

\end{center}

\vspace{.6cm}

\hrule \vspace{0.2cm}
\centerline{\small{\bf Abstract}}
\vspace{-0.2cm}
{\small\noindent We derive the finite-temperature equation of state of dark matter superfluids with 2-body and 3-body contact interactions. The latter case is relevant
to a recently proposed model of dark matter superfluidity that unifies the collisionless aspects of dark matter with the empirical success of MOdified Newtonian Dynamics
at fitting galactic rotation curves. The calculation uses a self-consistent mean-field approximation. It relies on the Hartree-Fock-Bogoliubov approximation and follows the Yukalov--Yukalova
proposal to circumvent the well-known Hohenberg-Martin dilemma. The resulting equation of state is consistent with a gapless spectrum, and simultaneously satisfies
the equation of motion for the condensate wavefunction. As an application, we derive the finite-temperature density profile for dark matter superfluids,
assuming spherical symmetry and uniform temperature. The density profiles consist of a nearly homogeneous superfluid core, surrounded by an isothermal ``atmosphere"
of normal particles, with the transition taking place at the critical density.} 
\vspace{0.3cm}
\noindent
\hrule
\def\thefootnote{\arabic{footnote}}
\setcounter{footnote}{0}

\section{Introduction}

There has been considerable interest in the possibility that dark matter (DM) forms a Bose-Einstein condensate (BEC) or a superfluid in galaxies.
This idea goes back to the 1990's~\cite{Sin:1992bg,Ji:1994xh} and has been studied in various contexts. In the first context, known
as BEC DM~\cite{Goodman:2000tg,Peebles:2000yy,Silverman:2002qx,Arbey:2003sj,Boehmer:2007um,Lee:2008ux,Lee:2008jp,Harko:2011xw,RindlerDaller:2011kx,Slepian:2011ev,Dwornik:2013fra,Guzman:2013rua,Harko:2014vya}, superfluidity is achieved through 2-body contact interactions, which are necessarily repulsive to achieve stability. The zero-temperature, mean-field equation of state is $P\sim n^2$, where $n$ is the DM number density.

A second context is ultra-light DM~\cite{Hu:2000ke,Hui:2016ltb,Fan:2016rda}, where DM is comprised of ultra-light axions with mass $\sim 10^{-22}$~eV. 
In this case, stability is achieved through quantum mechanical pressure, {\it i.e.}, the uncertainty principle. In the context of the QCD axion, it has
been argued that Bose-Einstein condensation can occur in galaxies through gravitational interactions~\cite{Sikivie:2009qn,Erken:2011dz}, though
its stability has been disputed~\cite{Guth:2014hsa}. Ultra-light and QCD axions share the property of very weak interactions. The occupation number is
large, resulting in a BEC, but, as a result of weak interactions, the dispersion relation is not linear at low momentum, {\it i.e.}, the low-energy excitations
are not phonons.

More recently, DM superfluidity has been proposed~\cite{Berezhiani:2015pia,Berezhiani:2015bqa,Berezhiani:2017tth} as a natural framework to realize the phenomenon of MOdified Newtonian Dynamics (MOND)~\cite{Milgrom:1983ca,Milgrom:1983pn,Bekenstein:1984tv}, an empirical acceleration law that has proven quite successful at matching detailed rotation curves of disk galaxies over many decades in mass and luminosity. Let $a_{\rm N}$ denote the usual Newtonian gravitational field generated from the observed distribution of baryonic matter alone, and $a_0\sim 10^{-8}{\rm cm}/{\rm s}^{2}$ denote the MOND critical acceleration. The MOND empirical law states that the total gravitational acceleration $a$ is approximately $a_{\rm N}$ in the regime $a_{\rm N}\gg a_0$, and approaches the geometric mean $\sqrt{a_{\rm N} a_0}$ whenever $a_{\rm N}\ll a_0$. See~\cite{Famaey:2011kh} for a review. 

The idea of~\cite{Berezhiani:2015pia,Berezhiani:2015bqa,Berezhiani:2017tth} is that the MOND acceleration law results from DM phonons,
which couple to ordinary matter and thus mediate a long-range force. See also~\cite{Khoury:2016egg,Khoury:2016ehj,Hodson:2016rck,Addazi:2018ivg,Cai:2017buj,Alexander:2018fjp,Hossenfelder:2018iym}. Thus the collisionless nature of DM and ``modified gravity" aspects of MOND are unified, as different manifestations of the same underlying substance. 
(See~\cite{Khoury:2014tka,Famaey:2017xou} for other recent proposals to obtain the MOND empirical law through the properties of DM.) The DM candidate in this case consists of axion-like particles with sufficiently strong self-interactions such that they thermalize in galaxies.\footnote{The DM particles must necessarily have microscopic interactions with baryons in order to general
phonon-baryon interactions in the superfluid phase, though the form of such microscopic interactions is model-dependent~\cite{Berezhiani:2015pia,Berezhiani:2015bqa}.} 
With $m\sim {\rm eV}$, the de Broglie wavelengths of DM particles overlap in the (cold and dense enough) central region of galaxies, resulting in Bose-Einstein condensation into a superfluid phase. 

The MOND force law requires a specific equation of state for the DM superfluid, $P \sim n^3$. For this particular choice, the resulting phonon effective Lagrangian is similar to the MOND scalar field theory~\cite{Bekenstein:1984tv}. Remarkably, this phonon effective theory is strikingly similar to that of the Unitary Fermi Gas ({\it e.g.},~\cite{Giorgini:2008zz}), which has generated much excitement in the cold atom community in recent years. The desired $P \sim n^3$ equation of state results from a simple effective theory with 3-body contact interactions.
Unfortunately the phonon Lagrangian in this case has a wrong sign compared to Bekenstein-Milgrom~\cite{Bekenstein:1984tv}, and hence can only serve as a toy model for DM superfluidity~\cite{Berezhiani:2015pia,Berezhiani:2015bqa}. We shall henceforth refer to this superfluid as the `3-body' case, to distinguish it from BEC DM with 2-body interactions (the `2-body' case).

An important phenomenological question, relevant to both 2-body and 3-body superfluids, is to determine the nature of the density profiles in galaxies. In the 3-body case,
the density profile was derived in~\cite{Berezhiani:2015pia,Berezhiani:2015bqa} using a mean-field Gross-Pitaevskii approach. This amounts to
assuming that DM is entirely in the condensed phase, an approximation valid only at zero temperatures and for weak interactions. A similar exercise was performed in the
2-body case in~\cite{Boehmer:2007um}. In both cases the density profile is determined by hydrostatic equilibrium, which for
spherical symmetry reduces to a Lane-Emden equation. This traces back to the polytropic form of the equation of state, $P\sim n^\alpha$. The resulting profile is cored,
with nearly homogeneous density, and terminates at a certain radius set by the strength of interactions. In a recent paper~\cite{Berezhiani:2017tth}, the cored
superfluid profile was matched to a collisionless (Navarro-Frenk-White~\cite{Navarro:1996gj}) envelope at a radius where the density is too low to maintain thermal equilibrium. 

In reality, owing to their velocity dispersion DM particles have a small, non-zero temperature in galaxies. According to Landau's phenomenological model,
a superfluid at finite (sub-critical) temperature behaves as a mixture of two fluids~\cite{Landau:1941vsj}: an inviscid superfluid component, and a ``normal" component,
which is viscous and carries entropy. The normal fluid is comprised of a gas of excitations, which at low temperature are phonons.  

In this paper we derive the finite-temperature DM density profile, for superfluids with 2-body and 3-body interactions. For this purpose we calculate the
finite-temperature equation of state $P = P(n,T)$ in each case. After briefly reviewing the ideal Bose gas (Sec.~\ref{ideal bose}),
we begin in Sec.~\ref{2body standard} with the case of superfluidity with 2-body interactions, following the standard Hartree-Fock textbook
derivation, {\it e.g.},~\cite{Brown:1992db}. In the process we encounter two technical problems, well-known in the statistical physics literature~\cite{YUKALOV2008461}:

\begin{itemize}

\item The first problem pertains to the fluctuation-corrected expression for the condensate wavefunction, $\Psi$. Over a certain range of chemical potential, 
the correction to $|\Psi|^2$ becomes complex. This is analogous to the situation in quantum field theory for a scalar field exhibiting spontaneous symmetry
breaking~\cite{Coleman:1973jx,Halperin:1973jh}. The one-loop correction gives a complex result if the potential is concave, which can be interpreted as an instability
of the system~\cite{Weinberg:1987vp}. In our case, this would manifest itself as a complex chemical potential. The concept of a complex chemical potential has been
explored in~\cite{Cragg:2005}, where it was related to the decay rate of the condensate.

\item A second problem with the standard analysis is the Hohenberg-Martin dilemma~\cite{HOHENBERG1965291}. By the Hugenholtz-Pines
theorem~\cite{PhysRev.116.489}, which is one of the Ward-Takahashi identities, the energy spectrum below $T_{\rm c}$ should be gapless:
$\lim_{\textbf{k} \rightarrow 0}\epsilon_{\bf{k}} = 0$. However, we find that this condition is incompatible with the equation of
motion for $|\Psi|^2$. 

\end{itemize}

To overcome both problems, we follow in Sec.~\ref{YYapproxn} the Yukalov--Yukalova proposal~\cite{YUKALOV2008461} and introduce two different chemical potentials --- one for the condensed phase, 
and another one for the normal phase. The two chemical potentials are distinct for $T \leq T_{\rm c}$, and allow us to simultaneously enforce the condition of a gapless spectrum and ensure that the self-consistency condition for the mean field is satisfied. The two chemical potentials become equal at the critical temperature. For $T \geq T_{\rm c}$, there is of course a single chemical potential,
associated with the conserved particle number. Our calculation includes the contribution from higher-order interaction terms, treated in the Hartree-Fock-Bogoliubov (HFB)
approximation, {\it e.g.}~\cite{Andersen:2003qj}. The result is cast in terms of the momentum distribution of particles $n_{\textbf{k}} = \langle  a_{\textbf{k}}^{\dagger} a_{\textbf{k}} \rangle$,
and the so-called anomalous averages $\sigma_{\textbf{k}} = \left\langle a_{\textbf{k}} a_{-\textbf{k}} \right\rangle$. While the anomalous averages are ignored in the oft-used Popov approximation~\cite{Popov1,Popov2}, we will see that keeping $\sigma_{\textbf{k}}$ is critical in overcoming the Hohenberg-Martin dilemma.

The 2-body equation of state, $P = P(n,T)$, is obtained in Sec.~\ref{2bodyeos} for both $T\leq T_{\rm c}$ and $T\geq T_{\rm c}$.
We also calculate the superfluid fraction in Sec.~\ref{sf fraction}, to be distinguished from the condensate fraction. The finite-temperature equation
of state of 2-body DM superfluidity was computed in an earlier paper by Slepian and Goodman~\cite{Slepian:2011ev}. Conceptually,
there are two major differences, detailed in Sec.~\ref{comp to Slepian}, between their calculation and ours: $i)$ unlike~\cite{Slepian:2011ev}, we
include the contribution from the anomalous averages; $ii)$ we use two separate chemical potentials. The difference between their equation of state and ours
is small in the dilute gas limit. For denser gases, however, the Slepian-Goodman equation of state displays ``an unphysical lobe" and becomes multi-valued~\cite{Slepian:2011ev}, whereas our
equation of state is well-behaved and single-valued throughout.

In Sec.~\ref{3body} we apply the Yukalov--Yukalova approach to a DM superfluid with 3-body interactions, the case of interest for
the MOND phenomenon~\cite{Berezhiani:2015pia,Berezhiani:2015bqa,Berezhiani:2017tth}. As before, we derive the equation
of state $P = P(n,T)$ for both $T\leq T_{\rm c}$ and $T\geq T_{\rm c}$, as well as the superfluid fraction.

With knowledge of the equation of state $P = P(n,T)$, we derive in Sec.~\ref{density profiles} the superfluid DM density profile in
both 2-body and 3-body cases, for a fiducial galaxy with mass $M \sim 10^9~{\rm M}_\odot$. For simplicity we focus on static, spherically-symmetric halos and ignore the contribution
of baryons. We also assume a constant temperature, $T = {\rm const.}$, consistent with thermal equilibrium. In particular, we ignore the transition to a collisionless profile, which is expected to occur at larger radius where the density is too low to sustain thermal equilibrium, as explored in~\cite{Berezhiani:2017tth}. We instead focus on the superfluid/normal
region, where thermal equilibrium is justified. 

The density profile is determined by the condition of hydrostatic equilibrium. It consists of a nearly homogeneous core, where the
superfluid component dominates, surrounded by an isothermal ``atmosphere", where the normal component
dominates. The phase transition from the superfluid core to the isothermal normal region occurs when the density drops to the
critical value. Our density profiles are broadly similar to those derived in~\cite{Berezhiani:2017tth} to fit detailed
galactic rotation curves. Our results form the basis of a more detailed investigation of DM density profiles
and explicit fits of galactic rotation curves, along the lines of~\cite{Berezhiani:2017tth}. When we include the phonon-mediated force,
we expect the rotation curves to closely reproduce the MOND phenomenology.

\section{Ideal Bose Gas}
\label{ideal bose}

To set the stage for our calculation, we begin with a brief review of the ideal Bose gas. A gas of particles
can be treated classically if it obeys the criterion~\cite{Pathria:1996hda}
\begin{equation}
n\lambda_{\rm th}^3 \equiv n\left(\frac{2\pi\hbar^2}{mk_{\rm B}T}\right)^{3/2}\ll 1 \, ,
\end{equation}
where $n$ is the number density of the gas, and $\lambda_{\rm th}$ is its thermal de Broglie wavelength. In this regime, the physical properties of the system can be expressed in terms of power series in this parameter, giving us an insight into the quantum effects. When the above criterion is violated and the thermal de Broglie wavelength of the gas is comparable to the inter-particle separation, then quantum-mechanical effects become significant, and the system can no longer be approximated as an ideal gas. When, at fixed density $n$, the temperature drops below the critical temperature,
\begin{equation}
T_{\rm c} = \frac{2\pi\hbar^2}{mk_{\rm B}}\left(\frac{n}{\zeta(3/2)}\right)^{2/3}\,,
\label{T_c ideal}
\end{equation}
where $\zeta$ denotes the Riemann zeta function, a macroscopically large number of particles occupy the ground state giving rise to a BEC. 
Conversely, at fixed temperature $T$, the phase transition occurs at a critical density 
\begin{equation}
n_{\rm c} = \zeta(3/2)\left(\frac{mk_{\rm B}T}{2\pi\hbar^2}\right)^{3/2}\,.
\label{n_c ideal}
\end{equation}
For $T < T_{\rm c}$, the fraction of particles in the condensate is given by  
\begin{equation}
\frac{N_{\rm cond}}{N} = 1- \left( \frac{T}{T_{\rm c}} \right)^{3/2}\,.
\end{equation}
The condensate fraction approaches unity as the temperature becomes zero. 

Meanwhile, the pressure, $P$, is given by
\begin{equation}
 P = 
\begin{dcases}
    \frac{k_{\rm B}T}{\lambda_{\rm th}^3}\zeta(5/2) \,; & T<T_{\rm c}\\
    nk_{\rm B}T\frac{\text{g}_{5/2}(z)}{\text{g}_{3/2}(z)}\,;   & T>T_{\rm c} \,,
\end{dcases}
\label{pressure_free}
\end{equation}
where the fugacity $z$ of the gas is related to the chemical potential $\mu$ by $z = e^{\mu/k_{\rm B}T}$, and $\text{g}_{\nu}(z) $ are the Bose-Einstein functions given by
\begin{equation}
\text{g}_{\nu}(z) = \frac{1}{\Gamma(\nu)}\int_{0}^{\infty}\frac{x^{\nu-1}\text{d}x}{z^{-1}e^x -1} = z + \frac{z^2}{2^{\nu}} + \frac{z^3}{3^{\nu}} + \ldots 
\end{equation}
As $T \rightarrow \infty$, we have $\mu \rightarrow - \infty$ such that $z\rightarrow 0$. In this regime the Bose-Einstein functions can be approximated by the linear term, $\text{g}_{\nu}(z) \simeq z$,
and the equation of state~\eqref{pressure_free} reduces to the usual ideal gas relation.

\section{2-Body Interactions: Standard Treatment}
\label{2body standard}

In this Section we begin by considering a Bose gas with 2-point contact interactions. This will serve as a warm-up for the relevant case of interest, {\it i.e.}, a system with 3-point contact interactions, 
which will be studied in Sec.~\ref{3body}. The derivation in this Section is standard and follows, {\it e.g.},~\cite{Brown:1992db}. 

For a system of particles in a box of volume $V$, the pressure is defined as the negative of the grand potential per unit volume, 
\begin{equation}
P =  -\frac{\Omega}{V} \,,
\end{equation}
where the grand potential is given by
\begin{equation}
\Omega = -k_{\rm B} T\,\text{ln}\, \left[\text{Tr }e^{-\beta \widetilde{H}}\right] \, ;\qquad \widetilde{H}\equiv \hat{H} - \mu \hat{N} \,,
\label{grandpotential}
\end{equation}
with $\beta \equiv \frac{1}{k_{\rm B}T}$. In terms of the bosonic field operator, $\psi(x)$, the Hamiltonian is 
\begin{equation}
\widetilde{H} = \hat{H} - \mu \hat{N} = \int \text{d}^3x \left[-\frac{\hbar^2}{2m} \psi^{\dagger}(x)\nabla^2 \psi(x) - \mu\psi^{\dagger}(x)\psi(x) + \frac{1}{2}g_2 \,\psi^{\dagger\, 2}(x) \psi^2(x)  \right]\, ,
\end{equation}
where $m$ is the particle mass, and $g_2$ is a parameter of dimensions ${\rm energy} \times {\rm volume}$ which controls the strength of the contact interactions. It is related to the $2\rightarrow 2$ scattering length via
\be
a =\frac{m g_2}{4\pi\hbar^2}\,.
\label{scat length}
\ee

We decompose $\psi(x)$ into two parts
\begin{equation}
\psi(x) = \Psi\, + \psi_1(x)\, ,
\label{fieldsplit}
\end{equation}
where $\Psi$ is the field operator for the condensate, and $\psi_1(x)$ is the operator for the fluctuations.
In terms of the usual ladder operators,
\begin{equation}
\Psi = \frac{a_0}{\sqrt{V}}\,;\qquad  \psi_1(x) = \frac{1}{\sqrt{V}} \sum_{\textbf{k}\neq 0 } a_{\textbf{k}}e^{{\rm i}\vec{k}\cdot \vec{x}}\,.
\label{ladder_expansion}
\end{equation}
At low temperatures, the ground state contains a macroscopically large number of particles with zero momentum. In this limit of large occupation number, we can
treat $a_0$ and $a^\dagger_0$ as commuting, classical variables~\cite{bogoliubovbook}. The operator $\Psi$ reduces to a condensate wavefunction, which sets the 
number density of particles with zero momentum: 
\be
n_0 = |\Psi|^2\,.
\label{n0def}
\ee
The number of zero-momentum particles is $N_0 = \int {\rm d}^3 x |\Psi|^2$.
Meanwhile, the fluctuation operator $\psi_1(x)$ is treated as a quantum field, whose excitations are $\vec{k}\neq 0$ particles. By construction,
$\langle \psi_1(x)\rangle = 0$. The number operator $\hat{N_1}$ for the fluctuations is
\be
\hat{N_1} =  \int \text{d}^3x\ \psi_1^{\dagger}(x)\psi_1(x) \  = \sum_{\textbf{k} \neq 0 } a^{\dagger}_{\textbf{k}}a_{\textbf{k}}  \,.
\ee
The number of particles in the normal phase is given by the expectation value $N_1 = \left\langle \hat{N_1} \right\rangle$.

In the mean-field approximation, we would expect all the particles to be in the condensate at zero temperature.
The mean-field Hamiltonian is
\begin{equation}
\widetilde{H}_{\rm MF} = \int \text{d}^3x\left[ -\mu|\Psi|^2 + \frac{1}{2}g_2|\Psi|^4 \right]\, .
\end{equation}
The equation of motion, $\frac{\delta \widetilde{H}_{\rm MF}}{\delta\Psi} = 0$, gives
\begin{equation}
|\Psi|^2  = \frac{\mu}{g_2}\qquad (\text{Mean-Field})\,.
\label{zero temp condensate}
\end{equation}
We see that repulsive interactions, $g_2>0$, implies a positive chemical potential for the condensate. This agrees with the intuitive notion of an increase in internal energy with increasing number of particles for a repulsive system. 

To find corrections to this result for low temperature and including fluctuations, we use a self-consistent mean-field approximation~\cite{chaikinlubensky}.
We begin by expanding the Hamiltonian in powers of the fluctuation field $\psi_1(x)$.  The first order terms can be ignored since $\left \langle \psi_1(x) \right\rangle = 0$. Restricting to second order terms, we find:
\begin{equation}
\widetilde{H} = \widetilde{H}_{\rm MF} + \sum_{\textbf{k}\neq 0 } \left(\hbar \omega_{\textbf{k}} a_{\textbf{k}}^{\dagger}a_{\textbf{k}} + \frac{1}{2}\Delta \left( a_{\textbf{k}}^{\dagger}a_{-\textbf{k}}^{\dagger}  + a_{-\textbf{k}}a_{\textbf{k}}   \right)\right)\,,
\end{equation}
where the dispersion relation is
\begin{equation}
\hbar \omega_{\textbf{k}}    = \frac{\hbar^2 k^2}{2m} + 2n_0g_2 - \mu \,,
\end{equation}
and we have introduced 
\begin{equation}
\Delta= n_0g_2 \,.
\end{equation}
Recall that $n_0 \equiv \left\langle |\Psi|^2 \right\rangle $ is the number density of condensed particles.

The off-diagonal terms in the second-order Hamiltonian can be eliminated as usual by a suitable Bogoliubov transformation, $a_{\textbf{k}}    = u_{\textbf{k}}   b_{\textbf{k}}    + v_{\textbf{-k}}   ^*b_{\textbf{-k}}   ^{\dagger}\,$. The Bogoliubov coefficients that cancel the off-diagonal terms are
\begin{equation}
u_{\textbf{k}}^2 = \frac{\hbar\omega_{\textbf{k}} + \epsilon_{\textbf{k}}}{2\epsilon_{\textbf{k}}} \,;\qquad \ v_{\textbf{k}}   ^2=\frac{\hbar\omega_{\textbf{k}}    - \epsilon_{\textbf{k}}   }{2\epsilon_{\textbf{k}}   }\, \,,
\label{Bogoliubov}
\end{equation}
where 
\begin{equation}
\epsilon_{\textbf{k}}    \equiv \sqrt{\hbar^2\omega_{\textbf{k}}   ^2-\Delta^2}=\sqrt{\left(\frac{\hbar^2k^2}{2m} + n_0g_2  -  \mu\right) \left(\frac{\hbar^2k^2}{2m} + 3n_0g_2 -\mu \right)}\, .
\label{epsilon_k}
\end{equation}
The Hamiltonian, up to quadratic order in fluctuations, becomes 
\begin{equation}
\widetilde{H} = V\left[ -\mu|\Psi|^2 + \frac{1}{2}g_2|\Psi|^4 
\right] + \sum_{{\textbf{k}}   \neq 0 } \left[ \epsilon_{\textbf{k}}   b_{\textbf{k}}   ^{\dagger}b_{\textbf{k}}    + \frac{1}{2}\left(\epsilon_{\textbf{k}}    - \hbar\omega_{\textbf{k}}    \right) \right]\, .
\label{Htildequad}
\end{equation}
The last term diverges linearly in three dimensions and can be regularized using dimensional regularization, as explained in Appendix A of~\cite{Andersen:2003qj}. In the continuum limit, the result is
\begin{equation}
 \frac{1}{2} \int \frac{{\rm d}^3k}{(2\pi)^3} \left(\epsilon_{\textbf{k}}    - \hbar\omega_{\textbf{k}}    \right) = \frac{8m^{3/2}\Delta^{5/2}}{15\pi^2\hbar^3} \,.
\label{dim reg}
\end{equation}
This result will be used repeatedly in the subsequent Sections, though the expression for $\Delta$ will change.

Since the Hamiltonian is diagonal in the new ladder operators, we expect the associated quasiparticles to follow the usual Bose-Einstein statistics,
\begin{equation}
\left\langle b_{\textbf{k}}   ^{\dagger}b_{\textbf{k}}   \right\rangle \, \, = \frac{1}{e^{\beta \epsilon_{\textbf{k}}}   -1}\,.
\label{quasiparticle distribution}
\end{equation}
This can also be derived explicitly by treating $b_{\textbf{k}}^{\dagger}b_{\textbf{k}}$ as the number operator for the quasiparticles and evaluating the contribution of this term in the grand potential. 

Using~\eqref{Htildequad}--\eqref{quasiparticle distribution}, we can calculate the grand potential $\Omega$ in~\eqref{grandpotential}. Going from the discrete case to the continuous limit, we obtain
\begin{equation}
\frac{\Omega}{V} = \left( -\mu  |\Psi|^2  + \frac{1}{2}g_2 |\Psi|^4 \right)  + k_{\rm B} T\int \frac{\text{d}^3k}{(2\pi)^3}\text{ln}\left(1-e^{-\beta\epsilon_{\textbf{k}}}\right)+ \frac{8m^{3/2}\Delta^{5/2}}{15\pi^2\hbar^3}\, .
\end{equation}
The equation of motion, $\frac{\delta \Omega}{\delta  |\Psi|^2 } = 0$, leads to the following implicit equation:
\begin{eqnarray}
\nonumber
|\Psi|^2 &=& \frac{\mu}{g_2}   -  \frac{4}{3\pi^2\hbar^3}(mg_2)^{3/2}|\Psi|^3 \\
\nonumber
 &-& \int\frac{\text{d}^3k}{(2\pi)^3}\frac{\frac{\hbar^2 k^2}{m} -2\mu + 3g_2|\Psi|^2}{\sqrt{\left(\frac{\hbar^2k^2}{2m}-\mu + g_2|\Psi|^2\right)\left(\frac{\hbar^2k^2}{2m}-\mu + 3g_2|\Psi|^2\right)}\left(e^{\beta \sqrt{\left(\frac{\hbar^2k^2}{2m}-\mu + g_2|\Psi|^2\right)\left(\frac{\hbar^2 k^2}{2m}-\mu + 3g_2|\Psi|^2\right)}}-1\right)}\,. \\ 
\label{Equation of motion}
\end{eqnarray}
This equation is the self-consistency condition for $|\Psi|^2$, including the effects of fluctuations. The first term on the right-hand side matches~\eqref{zero temp condensate}, while the last term goes to zero as $T\rightarrow 0$. However, even in this limit we see that there is a correction to~(\ref{zero temp condensate}) which increases as the interaction strength increases. This suggests that, at $T=0$, the presence of interactions results in the gas not completely condensing, because of fluctuations.

For $\frac{\mu}{3g_2}<   |\Psi|^2 < \frac{\mu}{g_2}$, the third term on the right-hand side is problematic because the term inside the square root is negative for small momenta. Writing the exponent as a power series, we see that this expression will only be real when the exponent is small enough that the power series can be accurately truncated at linear order. In general, we will also end up getting an imaginary component in the integral. This term is analogous to the one-loop correction carried out in particle physics for the effective potential of a scalar field that exhibits spontaneous symmetry breaking~\cite{Coleman:1973jx,Halperin:1973jh}.  The exact effective potential is real and convex, even though its perturbative expansion may have complex terms~\cite{Fujimoto:1982tc} since the one loop correction is proportional to the logarithm of the second derivative of the potential. Thus the one-loop correction gives a complex result if the potential is concave. In~\cite{Weinberg:1987vp}, the authors argue that the imaginary part arising from the perturbative series of the effective potential indicates an instability and relates to the decay rate~\cite{Cragg:2005}. In our case, this would correspond to a complex chemical potential~\cite{Cragg:2005} with its imaginary part giving8654 the decay rate of the condensate.

Instead, we would like to find a formalism that gives us a real perturbative effective potential. One approach to obtaining a real effective potential was explored in~\cite{Cahill:1993mg}, where the 
usual free energy $H + \int\text{d}^3x\, j\phi $ was modified to $H + \int  \text{d}^3x\,j\phi^2$, with $\phi$ being the scalar field and $j$ an external source. The idea is similar to that used in~\cite{Hawking:1982my}, where it was argued that the effective action should be seen as a function of $\left\langle\phi^2\right\rangle$ instead of $\left\langle\phi\right\rangle$. In the next Section, we will see that some intuition of this problem in our case of a BEC lends a solution similar to that of~\cite{Cahill:1993mg}. 

There is, however, another problem with the above analysis. According to the Hugenholtz-Pines theorem~\cite{PhysRev.116.489}, which is one of the Ward-Takahashi identities, the energy spectrum below $T_{\rm c}$ should be gapless:
\begin{equation}
\lim_{\textbf{k} \rightarrow 0}\epsilon_{\bf{k}} = 0\,;\qquad T\leq T_{\rm c}.
\label{Gapless}
\end{equation}
It is clear from~\eqref{epsilon_k} this condition can only be satisfied for $\mu = |\Psi|^2g_2$. On the other hand, if our solution obeys~(\ref{Equation of motion}), then the spectrum cannot be gapless. This problem is commonly termed the Hohenberg-Martin dilemma~\cite{HOHENBERG1965291}. Therefore, we must somehow find a way to ensure that the solution is both self-consistent, {\it i.e.}, it satisfies the equation of motion $\frac{\delta \Omega}{\delta  |\Psi|^2} = 0$, and is gapless, {\it i.e.}, it satisfies~(\ref{Gapless}). 

\section{2-Body Interactions: Improved Treatment}
\label{YYapproxn}

To overcome the Hohenberg-Martin dilemma, we follow the Yukalov--Yukalova proposal~\cite{YUKALOV2008461} and introduce two different chemical potentials --- one for the condensed phase, $\mu_0$, and another one for the normal phase, $\mu_1$. The two chemical potentials are distinct for $T \leq T_{\rm c}$, and allow us to simultaneously enforce the condition of a gapless spectrum while ensuring that the solution is conserving. As we will see the two chemical potentials become equal at $T = T_{\rm c}$. For $T \geq T_{\rm c}$, there is of course a single chemical potential, associated with the conserved particle number. 

Thus the grand potential~\eqref{grandpotential} for $T \leq T_{\rm c}$ is generalized as follows 
\begin{equation}
\Omega = -k_{\rm B}T\, \text{ln}\, \left[\text{Tr }e^{-\beta \widetilde{H}}\right] \, ;\qquad \widetilde{H}\equiv \hat{H} -\mu_0\hat{N_0} -\mu_1\hat{N_1} \,.
\end{equation}
As before we Fourier transform the field and express the Hamiltonian in terms of creation and annihilation operators. An important difference is that we will keep the contributions from higher-order terms as well. The result is
\begin{equation}
\widetilde{H} = \sum_{i=0}^4 \widetilde{H}^{(i)}\,,
\end{equation}
where
\begin{eqnarray}
\nonumber
\widetilde{H}^{(0)} &=&  \left( \frac{1}{2}n_0g_2 - \mu_0  \right)N_0\, ;\\ 
\nonumber
\widetilde{H}^{(1)} &=& 0 \,;\\
\nonumber
\widetilde{H}^{(2)} &=& \sum_{\textbf{k}\neq 0 }\left\{ \left(  \frac{\hbar^2k^2}{2m} + 2n_0g_2 - \mu_1  \right)a_{\textbf{k}}^{\dagger}a_{\textbf{k}} + \frac{1}{2}n_0g_2 \left( a_{\textbf{k}}^{\dagger}a_{-\textbf{k}}^{\dagger}  + a_{-\textbf{k}}a_{\textbf{k}}   \right)\right\}\, ; \\
\nonumber
\widetilde{H}^{(3)} &= & \sqrt{N_0}\frac{g_2}{V}  \sum_{\textbf{k}_i \neq 0}\left( a_{\textbf{k}_1}^{\dagger} a_{\textbf{k}_1+\textbf{k}_2} a_{-\textbf{k}_2} + a_{-\textbf{k}_2}^{\dagger}a_{\textbf{k}_1+\textbf{k}_2}^{\dagger}a_{\textbf{k}_1} \right)\, ; \\
\widetilde{H}^{(4)} &=& \frac{g_2}{2V}\sum_{\textbf{k}_i \neq 0}a_{\textbf{k}_1}^{\dagger}a_{\textbf{k}_2}^{\dagger}a_{\textbf{k}_2+\textbf{k}_3}a_{\textbf{k}_1-\textbf{k}_3}\, .
\label{H2body_expand}
\end{eqnarray}
The first-order term $\widetilde{H}^{(1)}$ vanishes once again because $\left \langle \psi_1(x) \right\rangle = 0$. 
Note that the dependence on the condensate wavefunction $\Psi$ is encoded in the number density $n_0 = |\Psi|^2 $. 

To simplify the third- and fourth-order terms, we use the HFB approximation, {\it e.g.}~\cite{Andersen:2003qj}. In this approximation the third-order contribution is neglected:
\begin{equation}
\widetilde{H}^{(3)}_{\rm HFB} = 0\,.
\end{equation}
To evaluate the $\widetilde{H}^{(4)}$ contribution, we introduce the momentum distribution of the particles 
\begin{equation}
 n_{\textbf{k}} \equiv  \langle  a_{\textbf{k}}^{\dagger} a_{\textbf{k}} \rangle\,,
 \end{equation}
 as well as the so-called anomalous average
\begin{equation}
\sigma_{\textbf{k}} \equiv \left\langle a_{\textbf{k}} a_{-\textbf{k}} \right\rangle\,.
\end{equation}
In the oft-used Popov approximation~\cite{Popov1,Popov2}, the anomalous averages are ignored. However, this is a reasonable
assumption only near the critical temperature. Instead in this Section we will see that 
keeping the anomalous average is essential in overcoming the Hohenberg-Martin dilemma.

Aside from the condensate number density, $n_0 = |\Psi|^2$, we introduce $n_1$ for the number density of particles in the normal phase.
The total number density is $n = n_0 + n_1$. Similarly, $\sigma$ will denote the density of the anomalous average.
In other words, in terms of $n_{\textbf{k}}$ and $\sigma_{\textbf{k}}$ we have
\begin{equation}
n_1 = \frac{1}{V}\sum_{{\textbf{k}} \neq 0}n_{\textbf{k}}\, ;\qquad \sigma = \frac{1}{V}\sum_{{\textbf{k}} \neq 0}\sigma_{\textbf{k}}\, .
\label{normal_anomalous_densities}
\end{equation}
The anomalous average is conventionally taken as real, which can be achieved by redefining $\psi_1(x)$ by a constant phase.

With this notation, the fourth-order term in the HFB approximation becomes 
\begin{equation}
\widetilde{H}^{(4)}_{\rm HFB} = C_4 + \frac{g_2}{2} \sum_{{\textbf{k}} \neq 0} \left( 4n_1 a_{\textbf{k}}^{\dagger}a_{\textbf{k}} +  \sigma\left(a_{\textbf{k}}^{\dagger}a_{-\textbf{k}}^{\dagger} +a_{-\textbf{k}}a_{\textbf{k}}\right) \right) \,.
\label{H42body}
\end{equation}
The constant $C_4$ has been added to ensure that the HFB approximation preserves the expectation value of the original Hamiltonian. That is, it is fixed by the requirement that 
$\left\langle \widetilde{H}^{(4)}\right\rangle = \left\langle \widetilde{H}^{(4)}_{\rm HFB}\right\rangle$. On the one hand, the expectation value of~\eqref{H42body} gives
\begin{equation}
\left\langle \widetilde{H}^{(4)}_{\rm HFB} \right\rangle   = C_4 + 2g_2V\left(n_1^2 +  \frac{1}{2}\sigma^2\right) \,.
\label{H4HFB_3}
\end{equation}
On the other hand, the last of~\eqref{H2body_expand} implies
\begin{equation}
\left\langle \widetilde{H}^{(4)}  \right\rangle = g_2V\left(n_1^2 +  \frac{1}{2}\sigma^2\right) \,.
\end{equation}
This fixes the constant to $C_4 = -g_2V \left(n_1^2 +  \frac{1}{2}\sigma^2\right)$. Therefore~\eqref{H42body} becomes
\begin{equation}
\widetilde{H}^{(4)}_{\rm HFB} = -g_2V \left(n_1^2 +  \frac{1}{2}\sigma^2\right) + \frac{g_2}{2} \sum_{{\textbf{k}} \neq 0} \left( 4n_1 a_{\textbf{k}}^{\dagger}a_{\textbf{k}} +  \sigma\left(a_{\textbf{k}}^{\dagger}a_{-\textbf{k}}^{\dagger} +a_{-\textbf{k}}a_{\textbf{k}}\right) \right)\,.
\label{H42body_final}
\end{equation}

Adding this to $\widetilde{H}^{(0)}$ and $\widetilde{H}^{(2)}$, the Hamiltonian in the HFB approximation becomes
\begin{equation}
\widetilde{H}_{\rm HFB} = E_0 + \sum_{{\textbf{k}} \neq 0 } \hbar \omega_{\textbf{k}}a_{\textbf{k}}^{\dagger}a_{\textbf{k}}  +\frac{1}{2}\sum_{{\textbf{k}} \neq 0 } \Delta_2 \left( a_{\textbf{k}}^{\dagger} a_{-\textbf{k}}^{\dagger} +a_{-\textbf{k}}a_{\textbf{k}}  \right)\,,
\label{HHFB2body}
\end{equation}
where, analogously to the previous Section, we have defined 
\begin{equation}
\hbar\omega_{\textbf{k}} \equiv \frac{\hbar^2k^2}{2m}  + 2ng_2  - \mu_1\, ; \qquad  \Delta_2 \equiv g_2\left( n_0 +\sigma\right)\, ; \qquad \epsilon_{\textbf{k}} =  \sqrt{\hbar^2\omega_{\textbf{k}}^2- \Delta_2^2}\, .
\label{2bodyquantities}
\end{equation}
The quantity $E_0$ is the zero-point energy, given by:
\begin{equation}
\frac{E_0}{V} \equiv \left( \frac{1}{2}n_0g_2 - \mu_0 \right)n_0 - g_2 \left(n_1^2  + \frac{1}{2}\sigma^2\right)\, .
\label{E0}
\end{equation}

We are now in a position to the fix the chemical potentials. The condensate chemical potential $\mu_0$ is fixed by the equation of motion, 
$\left\langle\frac{\partial \widetilde{H}_{\rm HFB} }{\partial \Psi}\right\rangle = 0$. This gives
\begin{equation}
\mu_0 = (2n - n_0 + \sigma)g_2\, .
\label{mu0}
\end{equation}
To fix $\mu_1$, we first perform a Bogoliubov transformation, ($a_{\textbf{k}}\, ,a_{\textbf{k}}^{\dagger}) \rightarrow (b_{\textbf{k}}\, ,b_{\textbf{k}}^{\dagger} $), similar to~(\ref{Bogoliubov}), to eliminate the off-diagonal terms. The resulting diagonal Hamiltonian is
\begin{equation}
\widetilde{H}_{\rm HFB} = E_0 + \frac{1}{2} \sum_{{\textbf{k}} \neq 0}(\epsilon_{\textbf{k}} - \hbar\omega_{\textbf{k}})  + \sum_{{\textbf{k}} \neq 0 }\epsilon_{\textbf{k}}b_{\textbf{k}}^{\dagger} b_{\textbf{k}}\, .
\label{Hdiagonal}
\end{equation}
The normal-phase chemical potential $\mu_1$ is fixed by demanding that the spectrum be gapless: $\underset{{\textbf{k}} \to 0}{\lim} \epsilon_{\textbf{k}} =0$. This gives
\begin{equation}
\mu_1 = (2n - n_0 - \sigma)g_2\, .
\label{mu1}
\end{equation}
Comparing (\ref{mu0}) and (\ref{mu1}), we find that in order to have two different chemical potentials, we cannot assume $\sigma=0$. Thus ignoring the anomalous averages would lead us back to the Hohenberg-Martin dilemma. 

We should stress that~\eqref{mu1} holds only for $T \leq T_{\rm c}$. For $T > T_{\rm c}$, the ground state is no longer accessible, and~(\ref{Gapless}) does not hold. Moreover, 
as $T \rightarrow T_{\rm c}$, the condensate fraction and the anomalous average both go zero, $n_0,\,\sigma \rightarrow 0$. As can be seen from~\eqref{mu0} and~\eqref{mu1},  
in this limit $\mu_0 = \mu_1 = 2ng_2$. This is consistent with the disappearance of the condensate and the two phases being replaced by a single phase of the normal Bose gas.
For $T \geq T_{\rm c}$, the system is described as usual by a single chemical potential $\mu$. We will come back to this point in Sec.~\ref{2bodyeos} when deriving the equation of state.

The momentum distribution of the the quasiparticles is given by the usual Bose-Einstein distribution and we get the same relation for $\langle b_{\textbf{k}}^{\dagger}b_{\textbf{k}}\rangle $ as in~(\ref{quasiparticle distribution}). Using this, it is straightforward to calculate the momentum distribution and anomalous average of normal-phase particles:
\begin{equation}
n_{\textbf{k}} = \langle  a_{\textbf{k}}^{\dagger} a_{\textbf{k}} \rangle = \frac{\hbar\omega_{\textbf{k}}}{2\epsilon_{\textbf{k}}}\text{coth}\left( \frac{\epsilon_{\textbf{k}}}{2k_{\rm B}T} \right) - \frac{1}{2}\, ; \qquad \sigma_{\textbf{k}} = \left\langle a_{\textbf{k}} a_{-\textbf{k}} \right\rangle = -\frac{\Delta_2}{2\epsilon_{\textbf{k}}}\text{coth}\left( \frac{\epsilon_{\textbf{k}}}{2k_{\rm B}T} \right)\, .
\label{normal_anomalous_average}
\end{equation}
Note that $\sigma_{\textbf{k}}$ is real, which traces back to the coefficients in the Bogoliubov transformation~\eqref{Bogoliubov} being real.
To proceed, it is convenient to define dimensionless variables, representing respectively the normal and anomalous fractions:
\begin{equation}
\eta = \frac{n_1}{n}\,;\qquad  \xi = \frac{\sigma}{n}\, .
\label{dimensionless}
\end{equation}
By definition, the condensate fraction is $\frac{n_0}{n} = 1- \eta$. Taking the continuum limit, we integrate $n_{\textbf{k}}$ and $\sigma_{\textbf{k}}$ to
obtain the normal and anomalous fractions:
\begin{eqnarray}
\nonumber
\eta &=&   \frac{s^{3}}{3\pi^2}\biggl( 1 +   \frac{3}{2\sqrt{2}} \int_0^{\infty}  \left( \sqrt{1+x^2}-1  \right)^{1/2}  \left[\text{coth}\left( \frac{s^2x}{2t}  \right) -1 \right]\text{d}x \biggr)\,;\\
\xi &=& \xi_0 -\frac{s^{3}}{2\sqrt{2}\pi^2} \int_0^{\infty}\frac{\left( \sqrt{1+x^2}-1  \right)^{1/2}}{\sqrt{1+x^2}}\left[\text{coth}\left( \frac{s^2x}{2t}  \right) -1 \right]\text{d}x \,,
\label{normal_anomalous_fraction}
\end{eqnarray}
where we have introduced
\begin{equation}
\gamma = \frac{mg_2n^{1/3}}{4\pi\hbar^2} = \left(a^3n\right)^{1/3}\,; \qquad s^2 = 4\pi \gamma (1-\eta + \xi)\,;\qquad  t = \frac{mk_{\rm B}T}{n^{2/3}\hbar^2}\, .
\label{gamma s t}
\end{equation}
Note that $\gamma^3$ measures the number of particles per scattering volume. Meanwhile, using~\eqref{T_c ideal} we see that $t \sim T/T_{\rm c}$. 
Equations~\eqref{normal_anomalous_fraction} are implicit expressions for $\eta$ and $\xi$. (For instance, the factors of $s$ on the right-hand side depend on $\xi$.) 
In what follows, we will solve these implicit relations numerically to obtain $\eta = \eta(n,T)$  and $\xi = \xi(n,T)$.

In the expression for $\xi$, the first term $\xi_0$ is divergent and arises from the use of contact interactions. It can be regularized using dimensional regularization, together with the limiting condition that the anomalous average vanishes as the interaction strength vanishes~\cite{PhysRevA.74.063623}. This gives $\xi_0 = \frac{2s^{2}}{\pi^{3/2}} \sqrt{\gamma (1-\eta)}$. However, as pointed out in~\cite{0953-4075-43-20-205302}, this result is problematic in that it does not lead to a second-order phase transition. To deal with this issue,~\cite{Yukalov2014} argued that the aforementioned expression for $\xi_0$ is valid only for weak interactions at small temperatures, while at high temperature $\xi_0$ should be set to zero. To take this into account, we propose to multiply $\xi_0$ by a phenomenological factor of $\left(1-\frac{T}{T_{\rm c}}\right)$, 
\begin{equation}
\xi_0 = \left(1-\frac{T}{T_{\rm c}}\right)\frac{2s^{2}}{\pi^{3/2}} \sqrt{\gamma (1-\eta)}\, ,
\label{dimreg}
\end{equation}
thereby ensuring that the anomalous average vanishes as the condensate disappears.

\subsection{Equation of state}
\label{2bodyeos}

We have everything at our disposal to calculate the equation of state of the gas, $P(n,T) = -\Omega/V$. Using the fact that $\langle b_{\textbf{k}}^{\dagger}b_{\textbf{k}} \rangle$ satisfies the Bose-Einstein distribution~\eqref{quasiparticle distribution}, and taking the continuum limit, we obtain
\begin{equation}
P(n,T) = -\frac{E_0}{V} - \frac{1}{2}\int \frac{\text{d}^3k}{(2\pi)^3} (\epsilon_{\textbf{k}} - \hbar \omega_{\textbf{k}}) - k_{\rm B}T\int \frac{\text{d}^3k}{(2\pi)^3}\,\text{ln}\left(1 - e^{-\beta\epsilon_{\textbf{k}}}\right)\, .
\label{P2body}
\end{equation}
We evaluate this for $T\leq T_{\rm c}$ and $T\geq T_{\rm c}$ separately.

\begin{itemize}

\item $T\leq T_{\rm c}$: The first term, proportional to the zero-point energy, can be evaluated by substituting~\eqref{mu0} into~\eqref{E0}:
\begin{equation}
\frac{E_0}{V} = -g_2n^2 \left( 1 + \xi^2 -\frac{1}{2}\left(1-\eta -\xi\right)^2\right) \,.
\end{equation}
The other terms can be expressed neatly in terms of $\Delta_2$, given in~\eqref{2bodyquantities}. In terms of dimensionless variables,
\begin{equation}
\Delta_2 =g_2 n (1 - \eta + \xi)\,.
\label{Delta2}
\end{equation}
The second term in~\eqref{P2body} can be evaluated using dimensional regularization. The result is identical to~\eqref{dim reg}, with $\Delta$ replaced by $\Delta_2$.
Meanwhile, the gapless condition that fixed $\mu_1$ to~\eqref{mu1} implies
\begin{equation}
\hbar\omega_{\textbf{k}} =  \frac{\hbar^2k^2}{2m} + \Delta_2 \,; \qquad  \epsilon_{\textbf{k}} = \sqrt{\frac{\hbar^2k^2}{2m}\left(\frac{\hbar^2k^2}{2m} + 2\Delta_2\right)}\,.
\label{omega_epsilon_simple}
\end{equation}
Note that $\lim_{\textbf{k}\rightarrow 0} \epsilon_{\textbf{k}} = 0$, as desired. Putting everything together, the equation of state reduces to
\begin{eqnarray}
\nonumber
P\left(n,T \leq T_{\rm c}\right) &= &g_2n^2\left( 1 + \xi^2 -\frac{1}{2}\left(1-\eta -\xi\right)^2\right) -  \frac{8m^{3/2}\Delta_2^{5/2} }{15\pi^2\hbar^3} \\
&&-~k_{\rm B}T\int \frac{\text{d}^3k}{(2\pi)^3}\text{ln}\left[1 - e^{-\beta\sqrt{\frac{\hbar^2k^2}{2m}\left(\frac{\hbar^2k^2}{2m} + 2\Delta_2 \right) }}\right]\,.
\label{condensed phase pressure}
\end{eqnarray}
The normal and anomalous fractions, $\eta = \eta(n,T)$  and $\xi = \xi(n,T)$, appearing in this equation will be obtained by numerically solving the implicit equations~\eqref{normal_anomalous_fraction}. 

\item $T\geq T_{\rm c}$:  By definition, above the critical temperature the condensate fraction and the anomalous average both vanish, $n_0 = \xi = 0$, hence $\eta = 1$. Furthermore, as discussed below~\eqref{mu1}, the normal Bose gas is described by a single chemical potential $\mu$.  In other words, in this regime
\begin{equation}
\Delta_2 =0\,;\qquad \epsilon_{\textbf{k}} = \hbar\omega_{\textbf{k}} =  \frac{\hbar^2k^2}{2m} + 2ng_2 - \mu \,;\qquad  \frac{E_0}{V} = - g_2n^2\,.
\end{equation}
The integral for the temperature-dependent term is now the usual Bose integral and can be written in terms of a polylogarithm.
The equation of state for $T\geq T_{\rm c}$ therefore becomes\footnote{The polylogarithm of order $n$ is defined as $\text{Li}_{n}(z) = \sum_{k =1}^\infty \frac{z^k}{k^n}$.}
\begin{equation}
P\left(n,T \geq T_{\rm c}\right) = g_2n^2 + \frac{\sqrt{2}\Gamma\left(5/2\right)}{3\pi^2\hbar^3}T^{5/2}m^{3/2}\text{Li}_{5/2}\left(e^{\beta(\mu - 2n g_2)}\right)\, .
\label{normal phase pressure}
\end{equation}
To find $\mu$, we first use the standard thermodynamics relation $\frac{\partial P}{\partial \mu} = n$:
\begin{equation}
n = \frac{\sqrt{2}\Gamma\left(5/2\right)}{3\pi^2\hbar^3}T^{3/2}m^{3/2}\text{Li}_{3/2}\left(e^{\beta(\mu - 2n g_2)}\right)\, .
\label{normal phase eqn}
\end{equation}
Inverting this equation gives $\mu = \mu(n,T)$, and substituting the result in~\eqref{normal phase pressure} yields the equation of state $P = P(n,T)$. 
Combined with~\eqref{condensed phase pressure}, we therefore have the entire pressure profile of the gas. 

\end{itemize}

It is interesting to consider the $T\rightarrow 0$ limit of~\eqref{condensed phase pressure}. In this limit, the normal fraction goes to zero, $\eta \rightarrow 0$, but the anomalous fraction $\xi$ remains finite. It is easy to show that the integral in the expression for $\xi$ in~\eqref{normal_anomalous_fraction} vanishes in this limit, hence $\xi \rightarrow \xi_0$, with~\eqref{dimreg} reducing to
\be
\xi_0 = \frac{8}{\sqrt{\pi}} \sqrt{a^3n} + \ldots
\ee
where we have used the dilute Bose gas limit, $a^3n\ll 1$. Meanwhile,~\eqref{Delta2} gives $\Delta_2 \simeq n g_2$.
Putting everything together, the pressure~\eqref{condensed phase pressure} in the $T\rightarrow 0$ limit is 
\be
P(T = 0) = \frac{2\pi\hbar^2a}{m}n^2 \left( 1 + \frac{112}{15\sqrt{\pi}} \sqrt{a^3n} + \ldots\right)\,.
\label{P02body}
\ee
This is the well-known dilute-gas expansion in powers of $\sqrt{a^3n}$~\cite{Lee:1957zza,Wu:1959zz,Sawada:1959zz,Hugenholtz:1959zz}. The leading term, $P\sim n^2$, agrees with the mean-field result. The corrections are due to fluctuations. 

\begin{figure}[h]
\centering   
\subfigure[Pressure versus density]{\label{fig:1a}\includegraphics[height=1.9in]{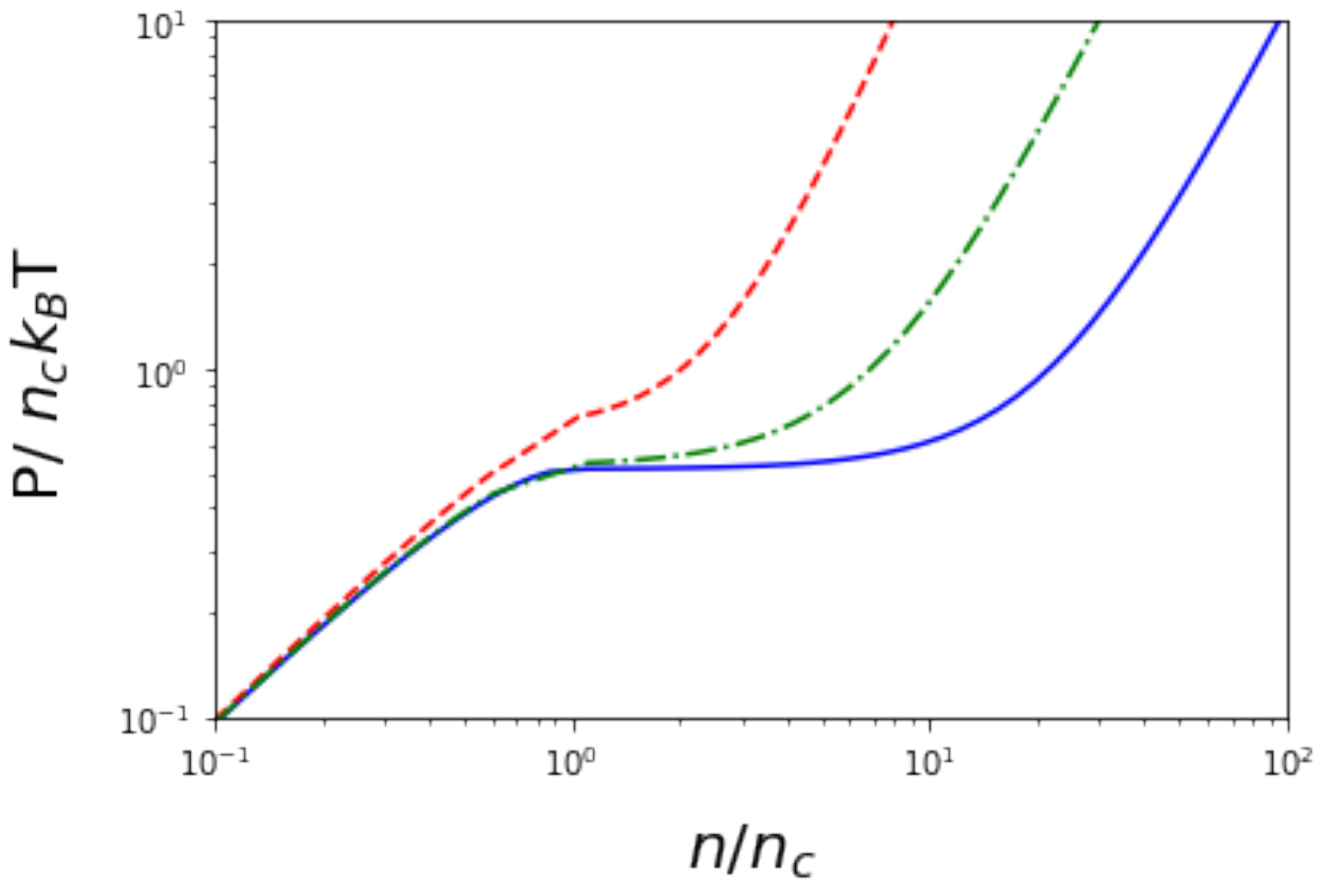}}
\subfigure[Pressure versus temperature]{\label{fig:1b}\includegraphics[height=1.9in]{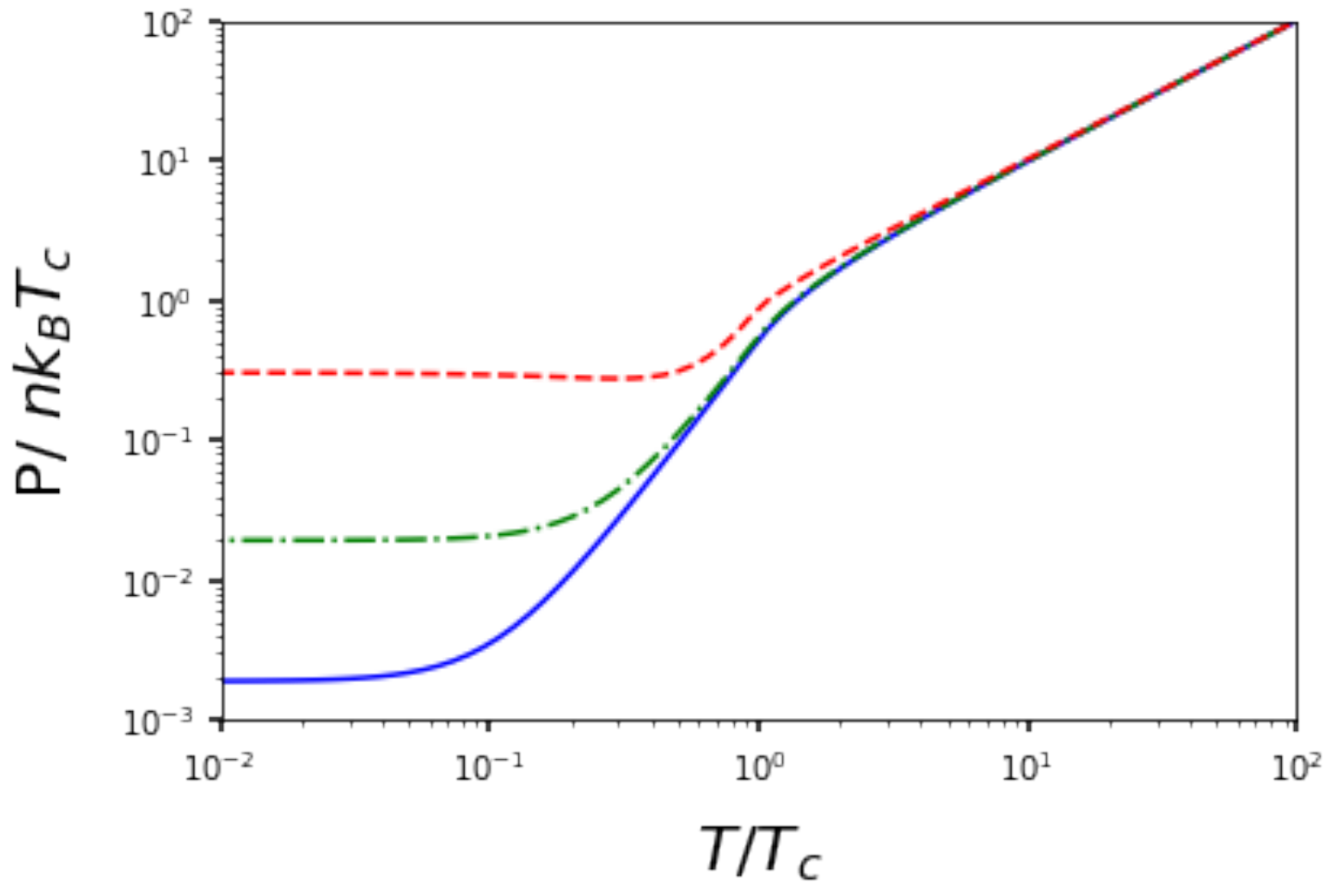}}
\caption{The equation of state for the 2-body case, given by~\eqref{condensed phase pressure} and~\eqref{normal phase pressure}, as a function of density (Left Panel) and temperature (Right Panel). Pressure is normalized by the ideal gas pressure at the critical point. Density and temperature are normalized by the respective quantities at the critical point. The red, green and blue curves correspond to $\gamma_c = 0.1$, 0.01 and 0.001 at the phase transition.}
\label{2body_eos}
\end{figure}

Figure~\ref{fig:1a} shows the pressure as a function of density. For small $n$, we get, $P \propto n$, which is reassuring since at low densities we would expect the system to behave like an ideal gas. For high densities, we get $P \propto n^2$, because of the contact interactions. Figure~\ref{fig:1b} shows the pressure as a function of temperature. We see that at large temperature, $P \propto T$ again confirming the ideal gas behavior. 

\subsection{Superfluid fraction}
\label{sf fraction}

At this stage it is important to distinguish between the condensate and the superfluid components. In general, the two phenomena are related to different aspects of the system. A BEC refers to the macroscopic occupation of the ground state. A superfluid, on the other hand, is a fluid whose long-wavelength excitations are phonons, {\it i.e.}, excitations with linear dispersion relation, $\omega_k \sim k$.  While superfluidity is related to strong pair correlation between particles, BEC relates to the coherence of the system. 

In this Section, we give a brief description of superfluidity and the superfluid fraction~\cite{Khalatnikov}. We begin by briefly reviewing Landau's criterion for superfluidity. In order for a fluid to exhibit superfluidity, its flow in a medium must be accompanied by zero friction, {\it i.e.}, no kinetic energy should be dissipated into heat. Since a quantum system heats up via discrete thermal excitations, we would like to find the condition under which the fluid cannot undergo a transition to the lowest energy excitation. 

Let the fluid be moving with a velocity $\textbf{v}$ at $T=0$, and the energy of these excitations be $\epsilon (\textbf{p})$ with associated momentum $\textbf{p}$. In the ``laboratory" reference frame, the energy of such an excitation would be 
\begin{equation}
E(\textbf{p}) = \epsilon(\textbf{p}) + \textbf{p}\cdot \textbf{v}\,.
\end{equation}
If $E(\textbf{p}) < 0$, then this excitation is energetically favorable. For $E(\textbf{p}) < 0$, the best case scenario is for the momentum of the excitation to be in the opposite direction to the fluid velocity. Thus, a necessary condition for the lowest energy excitation to occur is 
\begin{equation}
v > \text{min} \left(\frac{\epsilon(p)}{p} \right)\,.
\end{equation}
Since $v$ is positive definite, it follows that the fluid can exhibit superfluidity provided that 
\begin{equation}
\text{min} \left(\frac{\epsilon(p)}{p} \right) \neq 0\,.
\label{SFcond}
\end{equation}
This is Landau's criterion for superfluidity.

Since we did not make any assumption about the nature of the excitations,~\eqref{SFcond} is also valid at finite temperature. In that case, however, some excitations are present in the fluid {\it a priori}. These excitations can transfer energy to the walls of the medium, resulting in viscosity and normal fluid behavior. We therefore end up with a situation wherein, at finite temperature, part of the fluid behaves like a superfluid and moves without any viscosity, while the rest behaves like a normal fluid. This is Landau's phenomenological two-fluid picture. 

Let $\textbf{v}_{\rm s}$ and $\textbf{v}_{\rm n}$ be the velocities, and $\rho_{\rm s}$ and $\rho_{\rm n}$ the densities, of the superfluid and the normal components, respectively.
The distribution function for the elementary excitations depends on the relative motion between the superfluid and the normal components, and is characterized by 
\begin{equation}
E'(\textbf{p}) = E(\textbf{p}) - \textbf{p}\cdot \textbf{v}_{\rm n} = \epsilon(\textbf{p}) + \textbf{p}\cdot \textbf{v}_{\rm s} - \textbf{p}\cdot\textbf{v}_{\rm n}\,.
\end{equation}
The momentum density in the frame of the superfluid is given by 
\begin{equation}
\textbf{j}_0 = \rho_{\rm s}\textbf{v}_{\rm s} + \rho_{\rm n}\textbf{v}_{\rm n} - (\rho_{\rm s} + \rho_{\rm n}) \textbf{v}_{\rm s} = \rho_{\rm n}(\textbf{v}_{\rm n} - \textbf{v}_{\rm s})\,.
\label{superfluid_momentum_1}
\end{equation}
On the other hand, we can also calculate this expression in terms of the distribution function $n(\textbf{p})$ for phonons:
\begin{equation}
\textbf{j}_0 = \int \frac{\text{d}^3 p}{(2\pi\hbar)^3} \,\textbf{p}\,n(\textbf{p})\,.
\end{equation}
At low temperature, the relevant excitations are phonons. They have vanishing chemical potential, hence their distribution function is given by the Planck distribution:
$n(\textbf{p}) = \frac{1}{ e^{\beta E'(\textbf{p}) }-1}$. For small values of $\textbf{v}_{\rm n} - \textbf{v}_{\rm s}$, we Taylor expand $n(\textbf{p})$ in powers of the velocity difference. The zeroth-order term vanishes. Equating the first-order term with~\eqref{superfluid_momentum_1}, we obtain
\begin{equation}
\rho_{\rm n} = \frac{1}{3k_{\rm B}T}\int \frac{\text{d}^3p}{(2\pi\hbar)^3}p^2n' = \frac{1}{12mk_{\rm B}T}\int\frac{{\rm d}^3k}{(2\pi)^3} \frac{k^2}{\sinh^2\left(\frac{\epsilon_\textbf{k}}{2k_{\rm B}T}\right)} \,.
\end{equation}
Dividing this result by the total density and subtracting the result from unity gives the superfluid fraction~\cite{PhysRevA.76.013602} 
\begin{equation}
n_{\rm s} = 1 - \frac{s^5}{6\sqrt{2}\pi^2 t}\int_{0}^{\infty}\text{d}x \, \frac{x\,\left(\sqrt{1+x^2}-1 \right)^{3/2}}{\sqrt{1
+x^2}~\text{sinh}^2\left(\frac{s^2x}{2t}\right)}\, .
\label{Superfluid fraction}
\end{equation}

\begin{figure}[H]
\centering   
\subfigure[Condensate Fraction]{\label{fig:a}\includegraphics[height=2.1in]{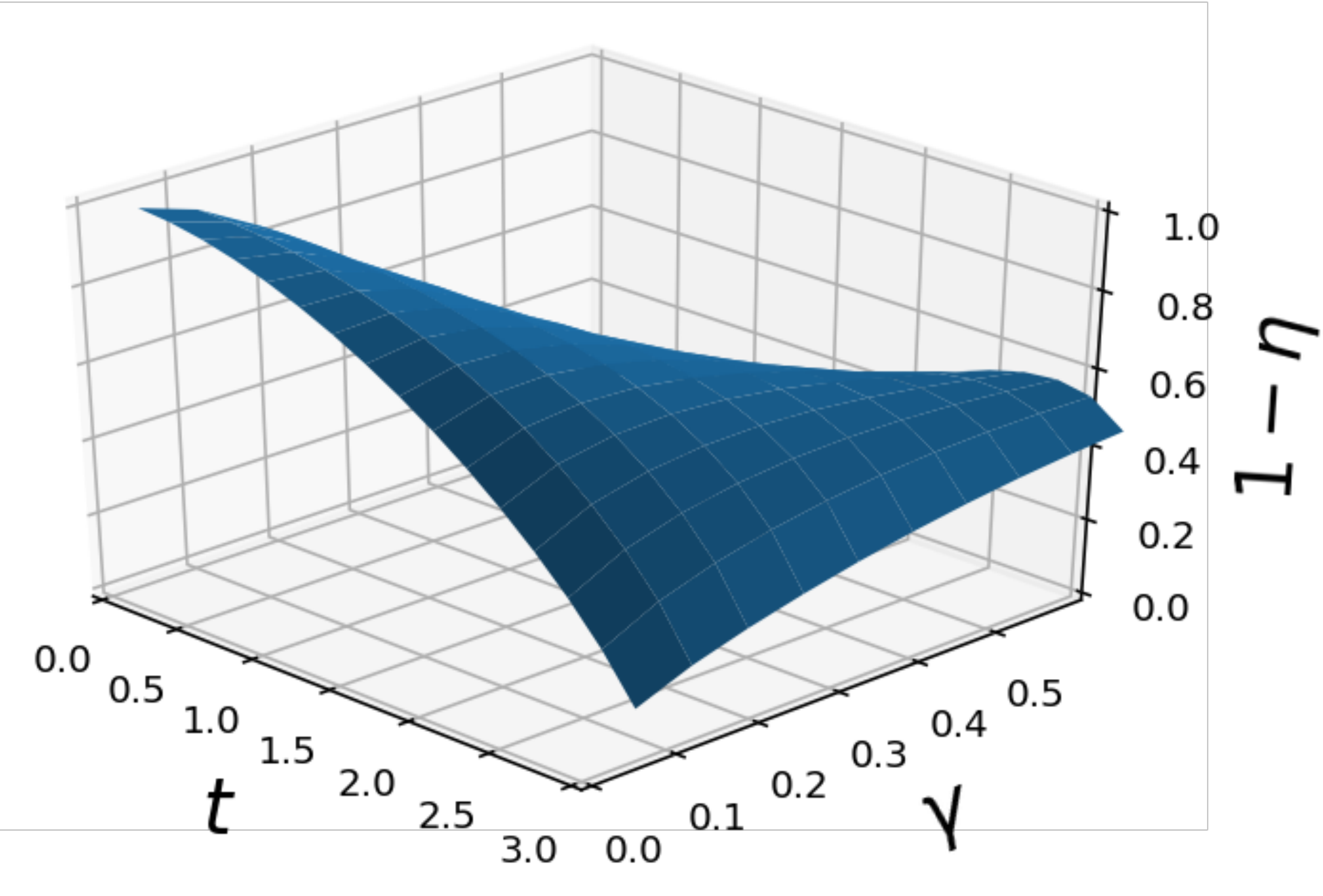}}
\subfigure[Superfluid Fraction]{\label{fig:b}\includegraphics[height=2.1in]{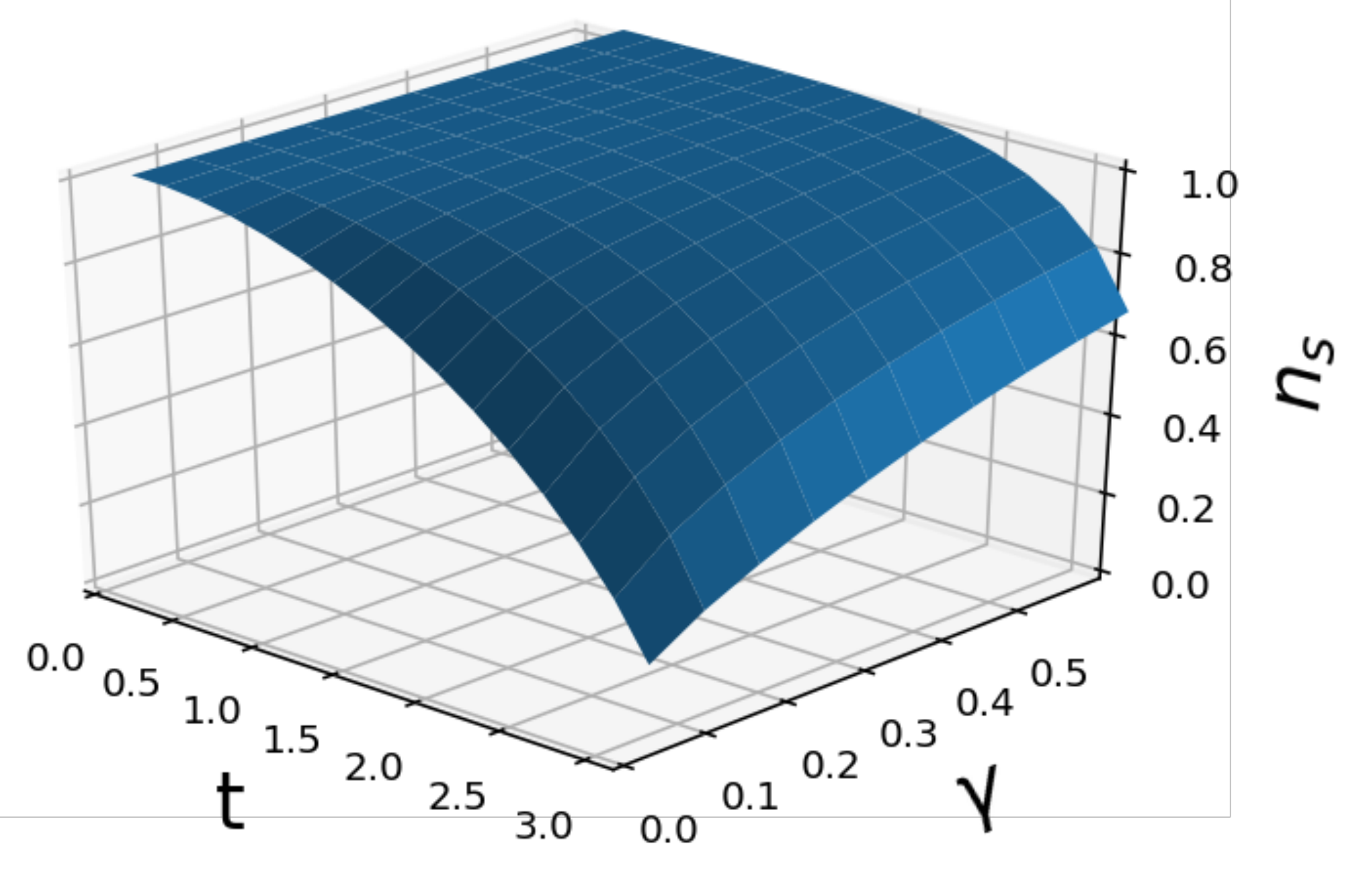}}
\caption{Comparing the difference between the condensate and the superfluid fractions.}
\label{2body condensate and superfluid fractions}
\end{figure}

Figure~\eqref{2body condensate and superfluid fractions} shows the condensate fraction (Left Panel) and superfluid fraction (Right Panel) as a function of the dimensionless temperature $t$ and 
interaction strength $\gamma$ defined in~\eqref{gamma s t}. In particular, note that at low temperature ($t\ll 1$) and sufficiently strong interaction ($\gamma \sim {\cal O}(1)$) the condensate fraction is very small whereas the superfluid fraction is close to unity. This highlights the difference between condensate and superfluid behavior. 

\subsection{Comparison to Slepian-Goodman}
\label{comp to Slepian}

We conclude this Section with a brief comparison to the paper by Slepian and Goodman~\cite{Slepian:2011ev}, who also computed the finite-temperature
equation of state for DM with 2-body interactions using the Hartree-Fock approximation. Conceptually, there are two major differences between their
calculation and ours: $i)$ we have included the contribution from the anomalous average, whereas~\cite{Slepian:2011ev} ignored it; $ii)$ we have used two
separate chemical potentials as described earlier. 
 
Instead of $\gamma$,~\cite{Slepian:2011ev} uses a different dimensionless variable,
\begin{equation}
\theta = \frac{g_2 n_{\rm c}}{k_{\rm B}T} = 2\big(\zeta(3/2)\big)^{2/3} \left(a^3n_{\rm c}\right)^{1/3}\,,
\label{theta}
\end{equation}
where in the last step we have used~\eqref{n_c ideal} to substitute for the critical density $n_{\rm c}$ at the given temperature $T$.
The dilute gas approximation is valid at the critical density for $\theta \ll 1$, corresponding to weak coupling. The relation to our
$\gamma$ is $\theta = 2\big(\zeta(3/2)\big)^{2/3}\gamma(n_{\rm c})$.

\begin{figure}[H]
     	\centering
        \includegraphics[height=3in]{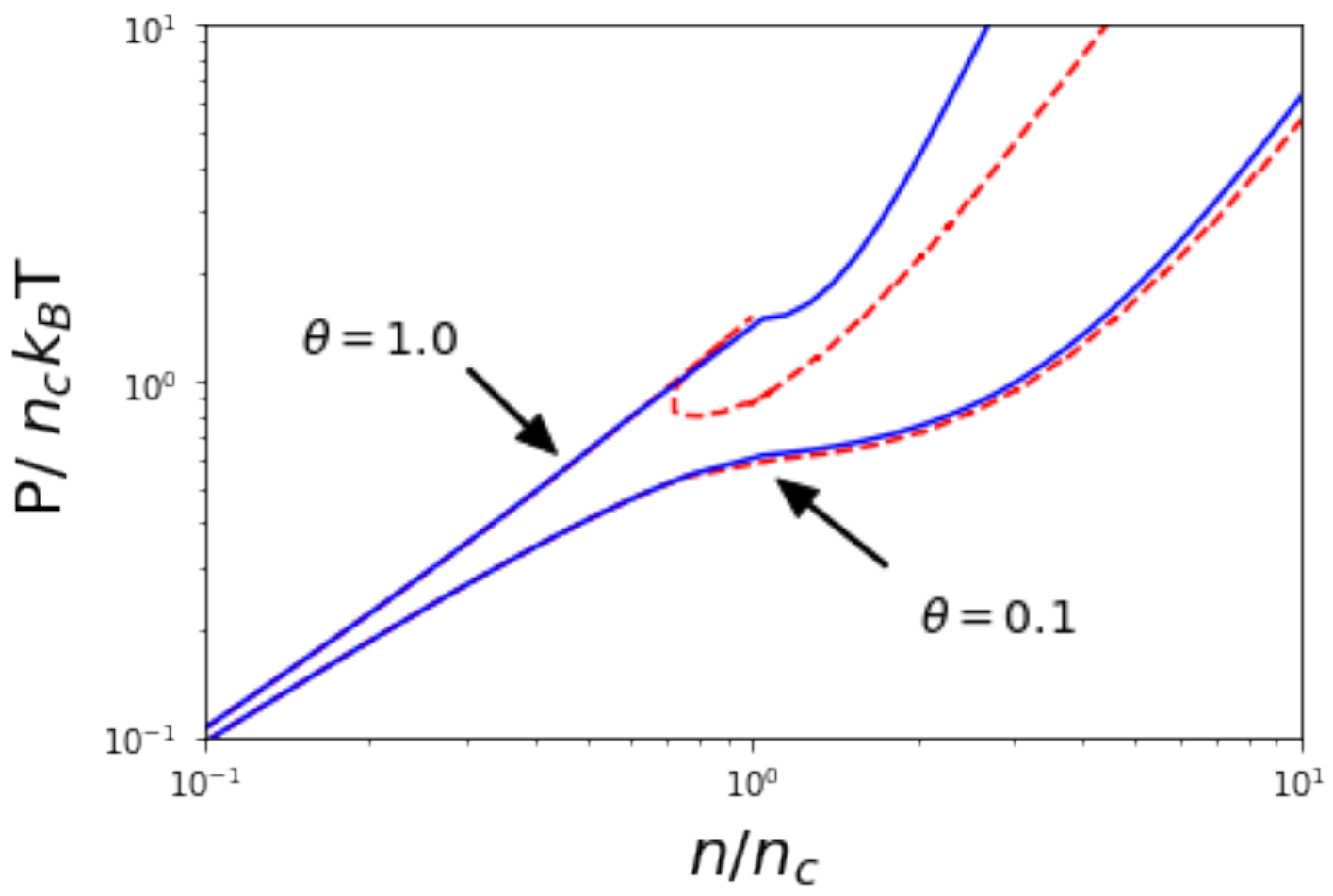}
    \caption{Comparison of our equation of state (solid blue), given by~\eqref{condensed phase pressure}, with that of Slepian and Goodman~\cite{Slepian:2011ev} (dashed red), given by~\eqref{Slepian_eos_3}. For $\theta = 0.1$ (bottom curves), corresponding to a relatively dilute gas, there is good agreement. Both curves behave as $P \propto n$ for $\frac{n}{n_{\rm c}} \ll 1$ and $P \propto n^2$ for $\frac{n}{n_{\rm c}} \gg 1$, as expected. For $\theta = 1$ (top curves), the Slepian-Goodman equation of state displays ``unphysical lobe" around $n = n_{\rm c}$ and becomes multi-valued, whereas our equation of state is well-behaved throughout.}
\label{SG}
\end{figure}

We briefly review the calculation of their equation of state, which is given by their Eqs.~(16)$-$(18). They begin
by defining the normal fraction as
\begin{equation}
\eta =   \frac{n_{\rm c}}{n} \frac{\text{Li}_{3/2}(z) }{\zeta(3/2)}\,.
\label{Slepian_eos_1}
\end{equation}
For $T > T_{\rm c}$, the normal fraction is set to unity, $\eta =1$, and~\eqref{Slepian_eos_1} can be used to solve for $z$. For $T \leq T_{\rm c}$,
on the other hand, they use
\begin{equation}
z = \text{exp}\left[-\theta\,\frac{n}{n_{\rm c}}\,(1-\eta)\right]\,.
\label{Slepian_eos_2}
\end{equation}
Substituting this into~\eqref{Slepian_eos_1} gives an implicit equation for $\eta$, which is the analogue of our~\eqref{normal_anomalous_fraction}. (Because~\cite{Slepian:2011ev} ignores the anomalous fraction $\sigma$, they obtain only one implicit equation instead of two.) The solution to this implicit equation gives the normal fraction $\eta$, which upon substitution in~\eqref{Slepian_eos_2} gives $z$. 
With the solution for $\eta$ and $z$ at hand, they obtain the equation of state\footnote{There is a minor typographical error in Eq.~(17) of~\cite{Slepian:2011ev}. Following the details in Appendix A of their paper, we found that $H(\hat{\nu}_0)$ should be replaced simply by $\hat{\nu}_0$. With this replacement, their Eq. (17) and our~\eqref{Slepian_eos_1} are consistent.} 
\begin{equation}
P = \left(1- \frac{1}{2}(1-\eta)^2 \right)g_2 n^2 + k_{\rm B} Tn_{\rm c}\zeta\left(\frac{3}{2}\right)^{-1}\text{Li}_{5/2}(z)\,.
\label{Slepian_eos_3}
\end{equation}

To show the behavior of their equation of state, they plot the pressure of the gas as a function of $n$ for different values of $\theta$. 
In Fig.~\ref{SG} we compare~\eqref{Slepian_eos_3} (dashed red curves) with our equation of state~\eqref{condensed phase pressure}
(solid blue curves) for $\theta = 0.1$  and for $\theta =1$. For $\theta = 0.1$, corresponding to a relatively dilute gas, there is good agreement. Both 
curves behave as expected, with $P \propto n$ for $\frac{n}{n_{\rm c}} \ll 1$ and $P \propto n^2$ for $\frac{n}{n_{\rm c}} \gg 1$, and the two regions 
are connected in a smooth manner. For $\theta = 1$, however, the Slepian-Goodman equation of state displays a pathological behavior for $n \approx n_{\rm c}$.
Instead of a smooth curve, there is a kink and ``an unphysical lobe" around $n = n_{\rm c}$, such that the pressure is not a unique function of density. 
In contrast, our equation of state shows no such pathology and remains single-valued throughout.

\section{3-Body Interactions}
\label{3body}

We now consider the case in which we have 3-body contact interactions:
\begin{equation}
\hat{H} = \int {\rm d}^3x \biggl[ -\frac{\hbar^2}{2m}\psi^{\dagger}(x)\nabla^2 \psi(x) + \frac{1}{3}g_3\, \psi^{\dagger\, 3}(x) \psi^3(x) \biggr] \, ,
\label{H3body initial}
\end{equation}
where the coupling constant $g_3$ has units of ${\rm energy}\times {\rm volume}^2$. As shown in~\cite{Berezhiani:2015pia,Berezhiani:2015bqa}, this theory results in an effective phonon Lagrangian with the desired mean-field, zero-temperature equation of state,
\be
P_{\rm MOND} = \frac{\hbar^6}{12\Lambda^2m^3} n^3\,.
\label{PMOND}
\ee
The relationship between $g_3$ and these parameters is
\be
g_3 = \frac{\hbar^6}{8\Lambda^2 m^3} \simeq 7.5 \times 10^3  \left(\frac{\Lambda}{10^{-3}~{\rm meV}}\right)^{-2} \left(\frac{mc^2}{{\rm 10\, eV}}\right)^{-3}~{\rm eV}\,\mu{\rm m}^6 \,.
\label{g3 Lambda}
\ee
In the analysis of~\cite{Berezhiani:2017tth}, the best-fit values were $mc^2 = 1~{\rm eV}$ and $\Lambda = 0.05~{\rm meV}$. We will be interested in somewhat different parameter values,
namely $mc^2 \sim 10~{\rm eV}$ and $\Lambda\sim 10^{-3}~{\rm meV}$, in order for the dilute gas approximation to be valid.

The effective phonon action that results from~\eqref{H3body initial} has the same power-law as the Bekenstein-Milgrom action~\cite{Bekenstein:1984tv}, ${\cal L} \sim \left((\vec{\nabla}\theta)^2\right)^{3/2}$. This shows that a non-analytic scalar Lagrangian arises from a superfluid medium whose Hamiltonian~\eqref{H3body initial} and equation of state~\eqref{PMOND}
are both analytic. With $g_3 > 0$, as required for stability, unfortunately the resulting phonon Lagrangian has a wrong sign compared to Bekenstein-Milgrom. The correct sign is
obtained for the unstable potential $g_3 < 0$, however in this case the interaction between bosons is attractive and hence the homogeneous BEC is unstable against collapse~\cite{Berezhiani:2015pia,Berezhiani:2015bqa}. Despite the sign difference with Bekenstein-Milgrom, the hexic model nevertheless serves as a toy model for DM superfluidity relevant for the MOND phenomenon~\cite{Berezhiani:2015pia,Berezhiani:2015bqa}. 

As before we introduce two chemical potentials and work with the shifted Hamiltonian $\tilde{H} = \hat{H} - \mu_0 \hat{N_0} - \mu_1\hat{N_1}$. 
We split the field operator as in~\eqref{fieldsplit}, and expand $\psi_1$ in terms of creation operators as in~\eqref{ladder_expansion}. 
Since $\widetilde{H}^{(1)}, \widetilde{H}^{(3)}$ and $\widetilde{H}^{(5)}$ are zero in the HFB approximation, we shall henceforth ignore them.
The remaining terms are:
\begin{eqnarray}
\nonumber
\widetilde{H}^{(0)} &=& \left(  -\mu_0 + \frac{g_3}{3}n_0^2 \right)N_0\, ;\\
\nonumber
\widetilde{H}^{(2)} &=& \sum_{{\textbf{k}} \neq 0} \left[ \left(  \frac{\hbar^2k^2}{2m} + 3g_3n_0^2 - \mu_1 \right)a_{\textbf{k}}^{\dagger}a_{\textbf{k}} + g_3n_0^2\left
(a_{\textbf{k}}^{\dagger}a_{-\textbf{k}}^{\dagger} + a_{-\textbf{k}}a_{\textbf{k}}   \right) \right]  \,;\\
\nonumber
\widetilde{H}^{(4)} &=& \frac{g_3n_0}{V}\sum_{\textbf{k}_i \neq 0} \left[ a_{\textbf{k}_1}^{\dagger}a_{\textbf{k}_2}^{\dagger}a_{\textbf{k}_3}^{\dagger}a_{\textbf{k}_1+\textbf{k}_2+\textbf{k}_3}  + a_{\textbf{k}_1+\textbf{k}_2+\textbf{k}_3}^{\dagger}a_{\textbf{k}_1}a_{\textbf{k}_2}a_{\textbf{k}_3} +  3a_{\textbf{k}_1}^{\dagger}a_{\textbf{k}_2}^{\dagger}a_{\textbf{k}_3}a_{\textbf{k}_1+\textbf{k}_2 -\textbf{k}_3}
\right]\, ;\\
\widetilde{H}^{(6)} &=& \frac{g_3}{3V^2}\sum_{\textbf{k}_i \neq 0}  a_{\textbf{k}_1}^{\dagger}a_{\textbf{k}_2}^{\dagger}a_{\textbf{k}_3}^{\dagger}a_{\textbf{k}_4}a_{\textbf{k}_5}a_{\textbf{k}_1+\textbf{k}_2+\textbf{k}_3-\textbf{k}_4-\textbf{k}_5} \, .
\label{H3body}
\end{eqnarray}
The dependence on the condensate wavefunction is encoded in the condensate number density, $n_0 = |\Psi|^2$. 

In the HFB approximation, the fourth-order contribution becomes:
\begin{eqnarray}
\nonumber
\widetilde{H}^{(4)}_{\rm HFB} &=& -6g_3N_0\left(n_1^2+ n_1\sigma + \frac{1}{2}\sigma^2  \right) \\
& & + 3g_3n_0\sum_{{\textbf{k}} \neq 0} \biggl[ 2(2n_1 + \sigma) a_{\textbf{k}}^{\dagger}a_{\textbf{k}}
+ (n_1 + \sigma) \left(a_{\textbf{k}}^{\dagger}a_{-\textbf{k}}^{\dagger}+ a_{-\textbf{k}}a_{\textbf{k}}\right)\biggr] \,,
\label{H4HFB}
\end{eqnarray}
where the normal and anomalous densities are given once again by~\eqref{normal_anomalous_densities}. The constant term was added to ensure that $\left\langle \widetilde{H}^{(4)}\right\rangle = \left\langle \widetilde{H}^{(4)}_{\rm HFB}\right\rangle$, similarly to the steps leading to~\eqref{H42body_final}. 

For the sixth-order term, the averaging procedure must be used twice. Averaging once gives
\begin{eqnarray}
\nonumber
\widetilde{H}^{(6)}_{\rm HFB} &= & \frac{g_3}{V^2} \sum_{\textbf{k}_i \neq  0} \bigg[3 a_{\textbf{k}_1}^{\dagger}a_{\textbf{k}_1}\left\langle a_{\textbf{k}_2}^{\dagger}a_{\textbf{k}_3}^{\dagger}a_{\textbf{k}_4}a_{\textbf{k}_2+\textbf{k}_3-\textbf{k}_4}  \right\rangle   +  a_{\textbf{k}_1}^{\dagger}a_{-\textbf{k}_1}^{\dagger}\left\langle a_{\textbf{k}_2+\textbf{k}_3+\textbf{k}_4}^{\dagger}a_{\textbf{k}_2}a_{\textbf{k}_3}a_{\textbf{k}_4}  \right\rangle  \\
&& \;\;\;\;\;\;\;\;\;\;\;\;\;\;\;+~ a_{-\textbf{k}_1}a_{\textbf{k}_1}\left\langle a_{\textbf{k}_2}^{\dagger}a_{\textbf{k}_3}^{\dagger} a_{\textbf{k}_4}^{\dagger}a_{\textbf{k}_2+\textbf{k}_3+\textbf{k}_4}\right\rangle \bigg] + C_6\, . 
\label{H6HFB_1}
\end{eqnarray}
The constant $C_6$ will be fixed shortly to ensure that the expectation of $\widetilde{H}^{(6)}_{\rm HFB}$ is the same as that of the original $\widetilde{H}^{(6)}$. The quartic correlators can be evaluated by averaging once more. For instance,
\begin{eqnarray}
\nonumber
\sum_{\textbf{k}_2,\textbf{k}_3,\textbf{k}_4 \neq  0} \left\langle a_{\textbf{k}_2}^{\dagger}a_{\textbf{k}_3}^{\dagger}a_{\textbf{k}_4}a_{\textbf{k}_2+\textbf{k}_3-\textbf{k}_4}  \right\rangle &=&  \sum_{\textbf{k}_2,\textbf{k}_3 \neq  0} \left(\langle a_{\textbf{k}_2}^{\dagger}a_{-\textbf{k}_2}^{\dagger}\rangle \langle a_{\textbf{k}_3}a_{-\textbf{k}_3}\rangle + 2 \langle a_{\textbf{k}_2}^{\dagger}a_{\textbf{k}_2}\rangle\langle a_{\textbf{k}_3}^{\dagger}a_{\textbf{k}_3}\rangle\right) \\
&=& V^2 \Big(2n_1^2 + \sigma^2\Big) \,.
\end{eqnarray}
Similarly, the other correlators give
\begin{equation}
\sum_{\textbf{k}_i \neq  0} \left\langle a_{\textbf{k}_2+\textbf{k}_3+\textbf{k}_4}^{\dagger}a_{\textbf{k}_2}a_{\textbf{k}_3}a_{\textbf{k}_4}  \right\rangle =
\sum_{\textbf{k}_i \neq  0} \left\langle a_{\textbf{k}_2}^{\dagger}a_{\textbf{k}_3}^{\dagger} a_{\textbf{k}_4}^{\dagger}a_{\textbf{k}_2+\textbf{k}_3+\textbf{k}_4}\right\rangle = 3V^2 n_1\sigma\,.
\end{equation}
The sixth-order Hamiltonian~\eqref{H6HFB_1} becomes
\begin{equation}
\widetilde{H}^{(6)}_{\rm HFB}  = 3g_3\sum_{\textbf{k} \neq 0 }\biggl[   \left(2n_1^2+\sigma^2\right) a_{\textbf{k}}^{\dagger}a_{\textbf{k}}  + n_1\sigma\left(a_{\textbf{k}}^{\dagger}a_{-\textbf{k}}^{\dagger}
+ a_{-\textbf{k}}a_{\textbf{k}}\right) \biggr] +C_6\, . 
\label{H6HFB_2}
\end{equation}
The constant $C_6$ is fixed by imposing $\left\langle \widetilde{H}^{(6)}\right\rangle = \left\langle \widetilde{H}^{(6)}_{\rm HFB}\right\rangle$. 
A straightforward calculation gives $C_6 = -2 g_3N_1 \left(2n_1^2 + 3\sigma^2\right)$. Substituting this into~\eqref{H6HFB_2}, we finally arrive at:
\begin{eqnarray}
\widetilde{H}^{(6)}_{\rm HFB} = - 2g_3N_1 \left(2n_1^2+ 3\sigma^2 \right)+ 3g_3\sum_{\textbf{k} \neq 0 }\biggl[   \left(2n_1^2+\sigma^2\right) a_{\textbf{k}}^{\dagger}a_{\textbf{k}}  + n_1\sigma\left(a_{\textbf{k}}^{\dagger}a_{-\textbf{k}}^{\dagger}
+ a_{-\textbf{k}}a_{\textbf{k}}\right) \biggr] \,.
\label{H6HFB}
\end{eqnarray}

Combining~\eqref{H4HFB} and~\eqref{H6HFB} with $\widetilde{H}^{(0)}$ and $\widetilde{H}^{(2)}$ from~\eqref{H3body}, the Hamiltonian in the HFB approximation once again takes the general form~\eqref{HHFB2body}
\begin{equation}
\widetilde{H}_{\rm HFB} = E_0 + \sum_{{\textbf{k}} \neq 0 } \hbar \omega_{\textbf{k}}a_{\textbf{k}}^{\dagger}a_{\textbf{k}}  +\frac{1}{2}\sum_{{\textbf{k}} \neq 0 } \Delta_3 \left( a_{\textbf{k}}^{\dagger} a_{-\textbf{k}}^{\dagger} +a_{-\textbf{k}}a_{\textbf{k}}  \right)\,,
\label{HHFB3body}
\end{equation}
where the zero-point energy is now given by
\begin{equation}
\frac{E_0}{V} = g_3\left\{ n_0\left(\frac{1}{3}n_0^2 - 6n_1\sigma -3\sigma^2 - 6n_1^2 - \frac{\mu_0}{g_3} \right) - 2n_1 \left(2n_1^2+ 3\sigma^2 \right)\right\} \,,
\label{E03body}
\end{equation}
and the coefficients of the quadratic terms by
\begin{eqnarray}
\nonumber
\hbar\omega_{\textbf{k}} &=&  \frac{\hbar^2k^2}{2m} + 3g_3\Big(n_0(n_0 + 2\sigma + 4n_1) + \sigma^2  + 2n_1^2 \Big) - \mu_1 \,;\\
\Delta_3 &=& 2g_3\left(n_0^2 + 3n_0n_1 + 3n\sigma\right)\,.
\label{omegaDelta3body}
\end{eqnarray}

At this point we can determine the chemical potentials. The chemical potential for the condensed phase is fixed by the equation of motion, 
$\left\langle\frac{\partial \widetilde{H}_{\rm HFB}}{\partial \Psi}\right\rangle = 0$, which gives
\begin{equation}
\mu_0 = g_3\Big(n_0^2+ 6nn_1 + 3\sigma^2 + 2 \sigma(2n + n_1)\Big)\,.
\label{mu03body}
\end{equation}
To determine the chemical potential for the normal phase, we must once again perform a Bogoliubov transformation to bring the Hamiltonian in the diagonal form~\eqref{Hdiagonal}.
The dispersion relation is 
\begin{equation}
\epsilon_{\textbf{k}} =  \sqrt{\hbar^2\omega_{\textbf{k}}^2- \Delta_3^2}\, ,
\end{equation}
with $\omega_{\textbf{k}}$ and $\Delta_3$ now given by~\eqref{omegaDelta3body}. Demanding that the spectrum be gapless, $\underset{{\textbf{k}} \to 0}{\lim} \epsilon_{\textbf{k}} =0$, fixes $\mu_1$ to
\begin{equation}
\mu_1 = g_3\Big(n_0^2 + 6nn_1 +3\sigma^2 -6n_1\sigma\Big)\,.
\label{mu13body}
\end{equation}

The normal and the anomalous averages, $n_{\textbf{k}}$ and $\sigma_{\textbf{k}}$, are given by the same relations~(\ref{normal_anomalous_average}). Integrating over momenta gives us the normal and anomalous fractions, $\eta = n_1/n$ and $\xi = \sigma/n$. The result is identical to~(\ref{normal_anomalous_fraction}):
\begin{eqnarray}
\nonumber
\eta &=&   \frac{s^{3}}{3\pi^2}\biggl( 1 +   \frac{3}{2\sqrt{2}} \int_0^{\infty}  \left( \sqrt{1+x^2}-1  \right)^{1/2}  \left[\text{coth}\left( \frac{s^2x}{2t}  \right) -1 \right]\text{d}x \biggr)\,;\\
\xi &=& \xi_0 -\frac{s^{3}}{2\sqrt{2}\pi^2} \int_0^{\infty}\frac{\left( \sqrt{1+x^2}-1  \right)^{1/2}}{\sqrt{1+x^2}}\left[\text{coth}\left( \frac{s^2x}{2t}  \right) -1 \right]\text{d}x \,,
\label{normal_anomalous_fraction_bis}
\end{eqnarray}
with $\gamma$, $s$ and $t$ now given by:
\begin{equation}
\gamma = \frac{g_3mn^{4/3}}{2\pi\hbar^2}\,;\qquad  s^2 = 4\pi\gamma \Big[(1-\eta)(1+2\eta)+3\xi\Big]\,;\qquad  t = \frac{mk_{\rm B}T}{n^{2/3}\hbar^2}\, .
\label{gammanew}
\end{equation}
The divergent term $\xi_0$ is calculated using dimensional regularization and multiplied by a phenomenological factor of $\left(1-\frac{T}{T_{\rm c}}\right)$.
The result is 

\begin{equation}
\xi_0 = \left(1-\frac{T}{T_{\rm c}}\right)\frac{2s^{2}}{\pi^{3/2}} \sqrt{\gamma (1-\eta)(1+ 2\eta)}\, .
\label{dimreg}
\end{equation}

\subsection{Equation of state}
\label{3bodyeos}

As before, in the continuum limit the pressure $P(n,T) = -\Omega/V$ is given by
\begin{equation}
P(n,T) = -\frac{E_0}{V} - \frac{1}{2}\int \frac{\text{d}^3k}{(2\pi)^3} (\epsilon_{\textbf{k}} - \hbar \omega_{\textbf{k}}) - k_{\rm B}T\int \frac{\text{d}^3k}{(2\pi)^3}\,\text{ln}\left(1 - e^{-\beta\epsilon_{\textbf{k}}}\right)\, .
\label{P3body}
\end{equation}
Once again we can be evaluated this expression for $T\leq T_{\rm c}$ and $T\geq T_{\rm c}$ separately.

\begin{itemize}

\item $T\leq T_{\rm c}$: The zero-point energy can be evaluated by substituting~\eqref{mu03body} into~\eqref{E03body}:
\begin{equation}
\frac{E_0}{V} = - 4 g_3n^3 \left\{ 1 + \xi^2\left(1+2\eta +\frac{2}{3}\xi\right)  +\frac{2}{3}(1-\eta-\xi)^3 -\frac{3}{2}(1-\eta-\xi)^2\right\}\,.
\end{equation}
The other terms can be expressed neatly in terms of $\Delta_3$, given in~\eqref{omegaDelta3body}. In terms of dimensionless variables, 
\begin{equation}
\Delta_3 = 2g_3n^2 \Big((1-\eta)(1+2\eta) +3\xi\Big) \,.
\label{Delta3}
\end{equation}
The second term in~\eqref{P3body}, evaluated using dimensional regularization, is identical to~\eqref{dim reg} with $\Delta$
replaced by $\Delta_3$. Analogously to~\eqref{omega_epsilon_simple}, the gapless condition that fixed $\mu_1$ to~\eqref{mu13body}
once again leads to 
\begin{equation}
\epsilon_{\textbf{k}} = \sqrt{\frac{\hbar^2k^2}{2m}\left(\frac{\hbar^2k^2}{2m} + 2\Delta_3\right)}\,.
\end{equation}
Putting everything together, the equation of state becomes
\begin{eqnarray}
\nonumber
P\left(n,T \leq T_{\rm c}\right) &= & 4 g_3n^3 \left\{ 1 + \xi^2\left(1+2\eta +\frac{2}{3}\xi\right)  +\frac{2}{3}(1-\eta-\xi)^3 -\frac{3}{2}(1-\eta-\xi)^2\right\}   \\
&&-~ \frac{8m^{3/2}\Delta_3^{5/2}}{15\pi^2\hbar^3 } -k_{\rm B}T\int \frac{\text{d}^3k}{(2\pi)^3}\text{ln}\left[1 - e^{-\beta\sqrt{\frac{\hbar^2k^2}{2m}\left(\frac{\hbar^2k^2}{2m} + 2\Delta_3 \right) }}\right]\,.
\label{condensed phase pressure 3body}
\end{eqnarray}
We will solve the implicit equations~\eqref{normal_anomalous_fraction_bis} numerically to obtain $\eta = \eta(n,T)$ and $\xi = \xi(n,T)$.

\item $T\geq T_{\rm c}$:  By definition, above the critical temperature the condensate fraction and the anomalous average both vanish, $n_0 = \xi = 0$, hence $\eta = 1$. Furthermore, as discussed below~\eqref{mu1}, the normal Bose gas is described by a single chemical potential $\mu$.  In other words, in this regime
\begin{equation}
\Delta_3 =0\,;\qquad \epsilon_{\textbf{k}} = \hbar\omega_{\textbf{k}} =  \frac{\hbar^2k^2}{2m} + 6g_3n^2 - \mu \,;\qquad  \frac{E_0}{V} = - 4g_3n^3\,.
\end{equation}
The integral for the temperature-dependent term is now the usual Bose integral and can be written in terms of a polylog as:
\begin{equation}
P\left(n,T \geq T_{\rm c}\right) = 4g_3n^3 + \frac{\sqrt{2}\Gamma\left(5/2\right)}{3\pi^2\hbar^3}T^{5/2}m^{3/2}\text{Li}_{5/2}\left(e^{\beta\left(\mu - 6n^2 g_3\right)}\right)\, .
\label{normal phase pressure 3body}
\end{equation}
To find $\mu$, we first use the standard thermodynamics relation $\frac{\partial P}{\partial \mu} = n$:
\begin{equation}
n = \frac{\sqrt{2}\Gamma\left(5/2\right)}{3\pi^2\hbar^3}T^{3/2}m^{3/2}\text{Li}_{3/2}\left(e^{\beta\left(\mu - 6n^2 g_3\right)}\right)\, .
\label{normal phase eqn}
\end{equation}
Inverting this equation gives $\mu = \mu(n,T)$, and substituting the result in~\eqref{normal phase pressure} yields the equation of state $P = P(n,T)$. 
Combined with~\eqref{condensed phase pressure 3body}, we therefore have the entire pressure profile of the gas. 

\end{itemize}

\begin{figure}[h]
\centering   
\subfigure[Pressure versus density]{\label{fig4:a}\includegraphics[height=1.9in]{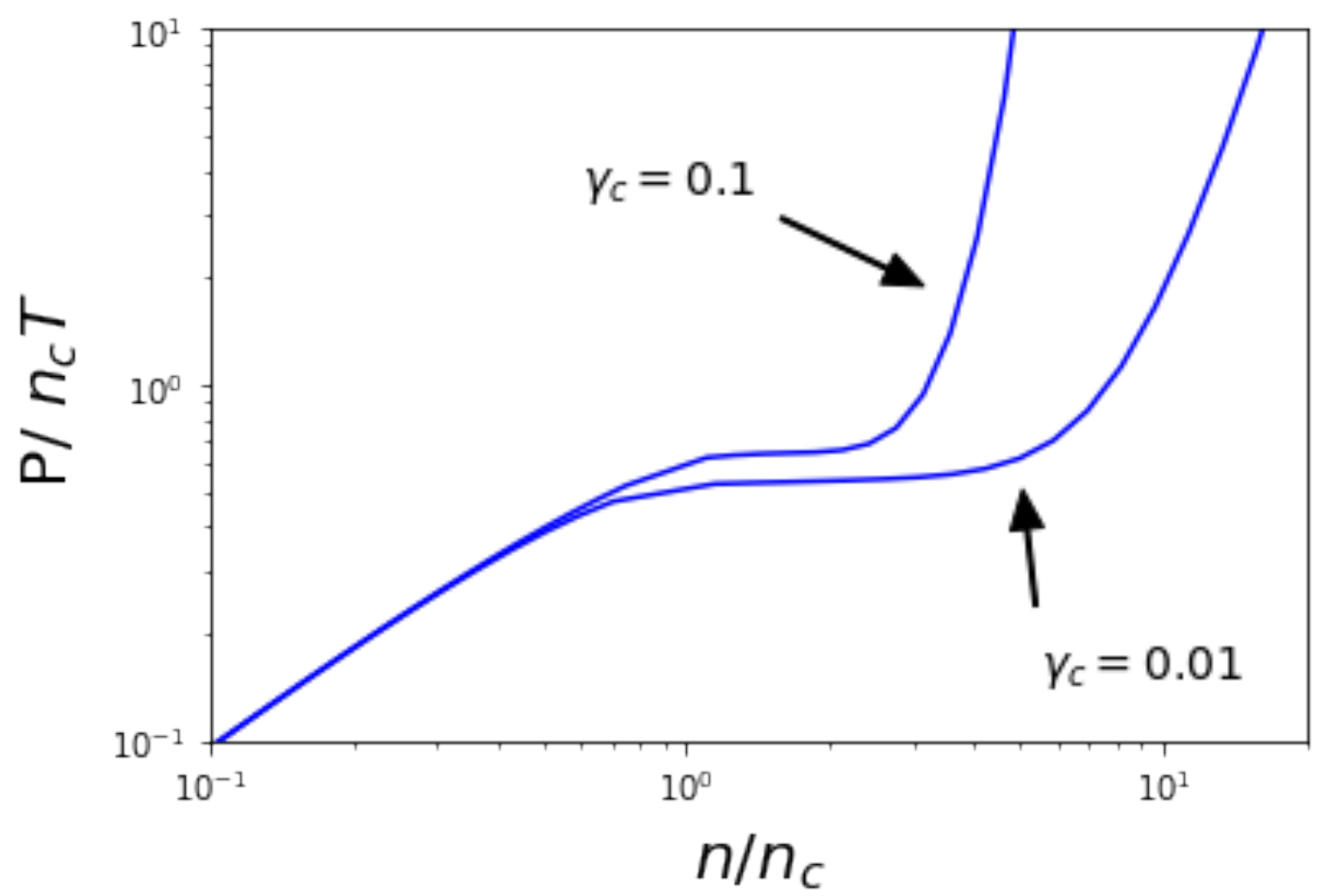}}
\subfigure[Pressure versus temperature]{\label{fig4:b}\includegraphics[height=1.9in]{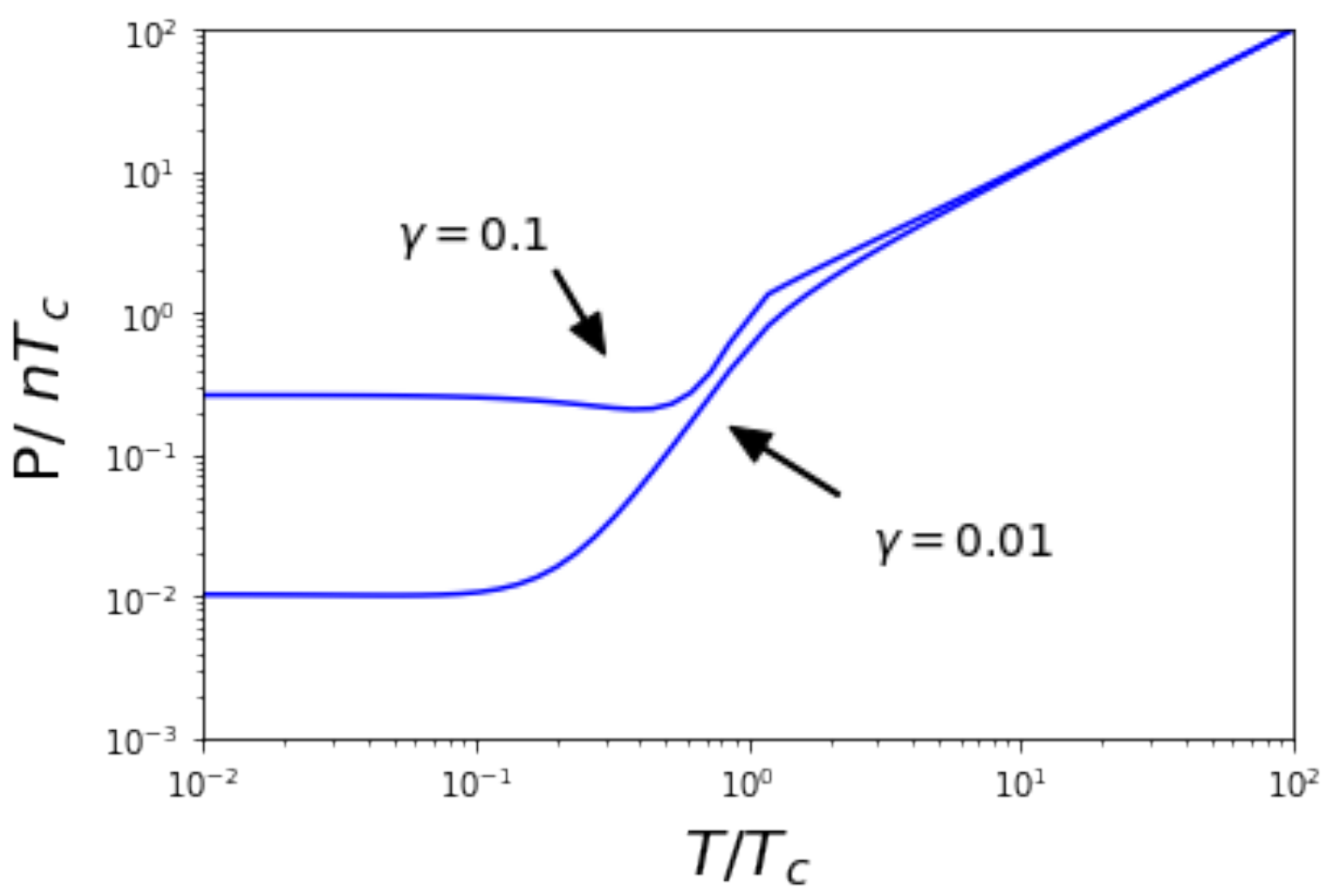}}
\caption{The equation of state for the 2-body case, given by~\eqref{condensed phase pressure 3body} and~\eqref{normal phase pressure 3body}, as a function of density (Left Panel) and temperature (Right Panel). Pressure is normalized by the ideal gas pressure at the critical point. Density and temperature are normalized by the respective quantities at the critical point.}
\end{figure}

Once again it is instructive to consider the $T\rightarrow 0$ limit. As in the 2-body case, the normal fraction goes to zero, $\eta \rightarrow 0$, but the anomalous fraction $\xi \simeq \xi_0$
remains finite. Using~\eqref{gammanew}, we have
\be
\xi_0 = \frac{1}{\pi^2} \left(\frac{2mg_3}{\hbar^2}\right)^{3/2} n^2 + \ldots
\ee
The dilute Bose gas limit therefore corresponds to
\be
\frac{1}{\pi} \left(\frac{mg_3}{\hbar^2}\right)^{3/4} n\ll 1\,,
\label{dilute 3body}
\ee
which is equivalent to $\gamma \ll 1$. Meanwhile,~\eqref{Delta3} gives $\Delta_3 \simeq 2g_3n^2$. Putting everything together, the zero-temperature equation of state is
\be
P(T = 0) = \frac{2}{3}g_3n^3 \left( 1 + \frac{22}{5\pi^2}\left(\frac{2mg_3}{\hbar^2}\right)^{3/2} n^2 + \ldots\right)\,.
\label{P03body}
\ee
This is the analogue of~\eqref{P02body} for 3-body interactions. The leading term, $P\sim n^3$, agrees with the mean-field result, while the corrections are due to fluctuations. 

Figure~\ref{fig4:a} shows the pressure as a function of density. We see that $P \propto n$ for small $n$, consistent with an ideal gas, whereas $P \propto n^3$ for large $n$.
Figure~\ref{fig4:b} shows the pressure as a function of temperature. We see that at large temperature, $P \propto T$ again confirming the ideal gas behavior. 

\begin{figure}[H]
\centering   
\subfigure[Condensate Fraction]{\label{fig:a}\includegraphics[height=2.1in]{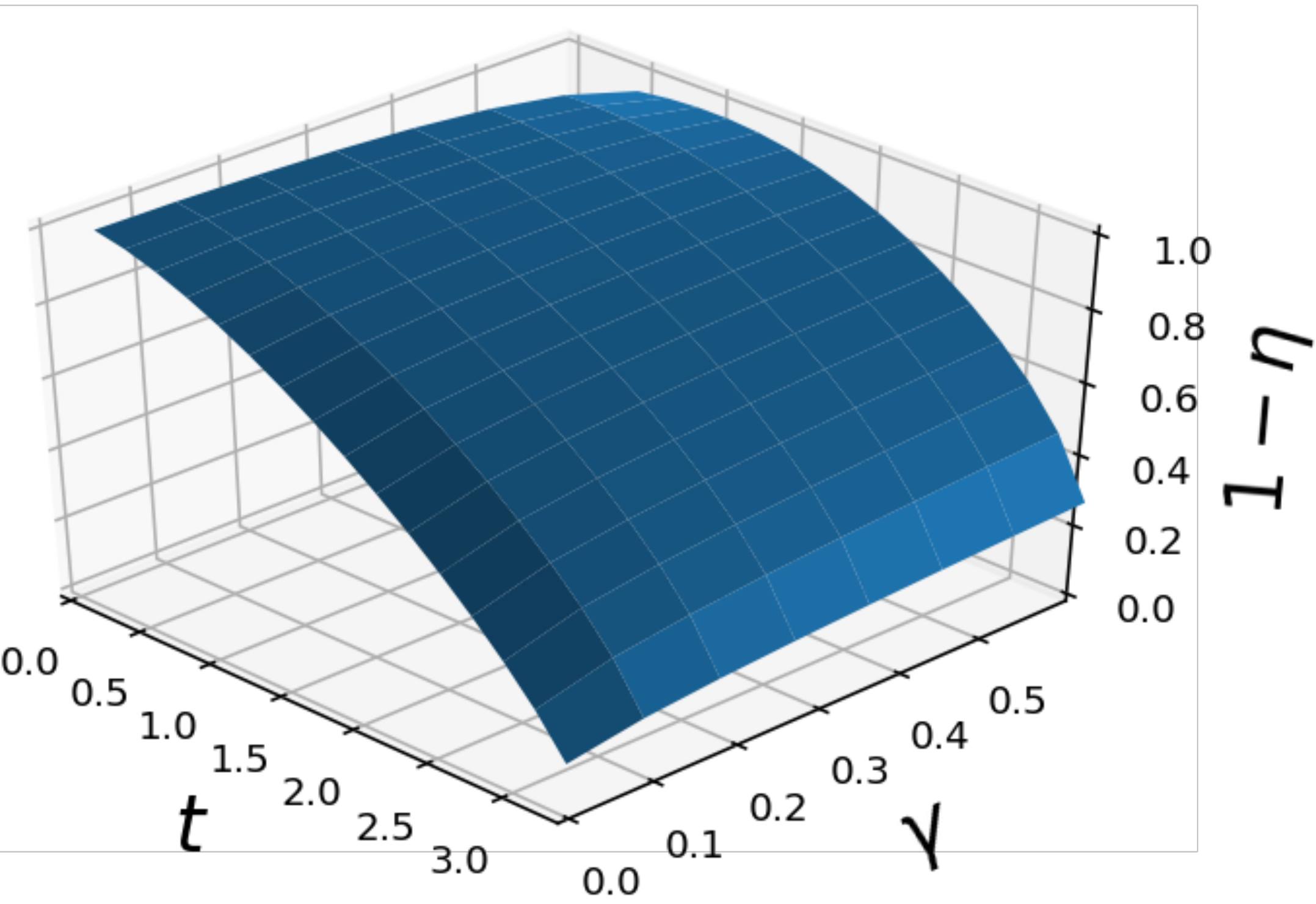}}
\subfigure[Superfluid Fraction]{\label{fig:b}\includegraphics[height=2.1in]{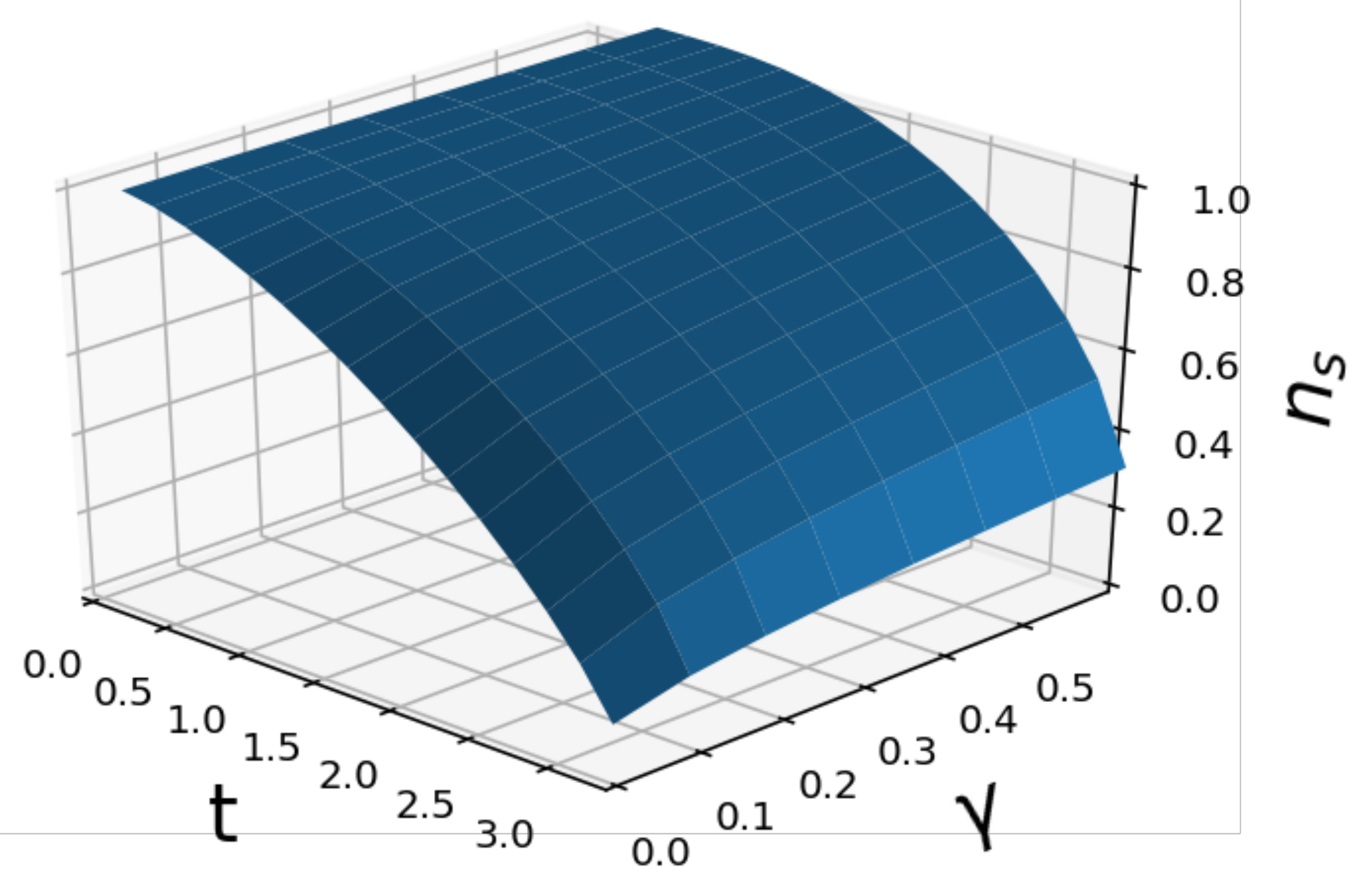}}
\caption{Comparison between the condensate fraction (Left Panel) and the superfluid fraction (Right Panel). For strongly interacting gases ($\gamma\sim {\cal O}(1)$, the superfluid fraction remains close to unity, despite the condensate fraction becoming small. The superfluid fraction is the relevant quantity to get the MOND phenomenon.}
\label{3pt_Sf}
\end{figure}

\subsection{Superfluid fraction}

The derivation of the superfluid fraction in Sec.~\ref{sf fraction} applies verbatim to the present case. The expression~\eqref{Superfluid fraction} for $n_{\rm s}$ is still valid, except that the dimensionless parameters are now given by~\eqref{gammanew}. Figure~\eqref{3pt_Sf} shows the condensate fraction (Left Panel) and superfluid fraction (Right Panel) as a function of the dimensionless temperature $t$ and interaction strength $\gamma$. In particular, we note that in the low temperature ($t\ll 1$) and strong interaction ($\gamma \sim {\cal O}(1)$) regime, the superfluid fraction remains close to unity, whereas the condensate fraction goes to zero. This corner of parameter space is relevant to the MOND phenomenon, and it is reassuring to see that the superfluid component dominates in this regime.

\section{Application: Finite-Temperature Density Profiles}
\label{density profiles}

With knowledge of the equation of state $P = P(n,T)$, we can now calculate the superfluid DM density profile. For this purpose we will make a number of simplifying
approximations, leaving to future work a more detailed derivation of density profiles and explicit fits of galactic rotation curves along the lines of~\cite{Berezhiani:2017tth}. 

Specifically, we focus on spherically-symmetric and static profiles, neglecting the contribution of baryons. 
The density profile is specified by the condition of hydrostatic equilibrium, $\frac{{\rm d}P(r)}{{\rm d}r} = -  mn(r)\frac{{\rm d}\Phi}{{\rm d}r}$, 
where the gravitational potential satisfies the Poisson equation, $\frac{1}{r^2} \frac{{\rm d}}{{\rm d}r}\left( r^2  \frac{{\rm d}\Phi}{{\rm d}r}\right) = 4\pi G_{\rm N}mn(r)$.
Combining these equations we obtain
\begin{equation}
\frac{1}{r^2}\frac{{\rm d}}{{\rm d}r}\left(  \frac{r^2}{n(r)}\frac{{\rm d} P(n,T)}{{\rm d}r} \right) = - 4\pi G_{\rm N}m^2n(r)\, ,
\label{hydrostatic_eqn}
\end{equation}
supplemented by smooth boundary conditions at the origin: $n(r=0) = n_0$ and $\frac{{\rm d}n}{{\rm d}r}(r=0) = 0$.

We assume that the superfluid has a constant temperature, $T = {\rm const.}$, consistent with thermal equilibrium. 
As we will see, the resulting density profile consists of a nearly homogeneous core, where the
superfluid component dominates, surrounded by an isothermal ``atmosphere", where the normal component
dominates. In the normal phase the equation of state is approximately that of an ideal gas, $P \simeq k_{\rm B}T mn$,
and the solution to~\eqref{hydrostatic_eqn} is the usual isothermal profile, $\rho \sim 1/r^2$, with the enclosed mass
increasingly linearly, $M(r) \sim r$. In this region the relation between temperature and enclosed mass is
$k_{\rm B} T = \frac{G_{\rm N}m M(r)}{2r} = {\rm constant}$.

Let us denote by $R_{\rm sf}$ denote the superfluid core radius at which the phase transition takes place. The core mass
is similarly denoted by $M_{\rm sf}$. By continuity, the temperature of the superfluid region can then be expressed as
\begin{equation}
T = \frac{G_{\rm N} M_{\rm sf}}{2R_{\rm sf}}\frac{m}{k_{\rm B}} = ~\frac{mc^2}{{\rm 10\, eV}}~\frac{M_{\rm sf}}{10^{9}~{\rm M}_\odot} \left(\frac{R_{\rm sf}}{10~{\rm kpc}}\right)^{-1}3 \times 10^{-4}~{\rm K} \,.
\label{haloT}
\end{equation}
Since the temperature is fixed, the phase transition occurs at a critical density $n_{\rm c}$ given by~\eqref{n_c ideal}:
\be
n_{\rm c} =  \left(\frac{mc^2}{{\rm 10\,eV}}\right)^3 \left(\frac{M_{\rm sf}}{10^{9}~{\rm M}_\odot}\right)^{3/2} \left(\frac{R_{\rm sf}}{10~{\rm kpc}}\right)^{-3/2}2.6\times10^3\,{\rm cm}^{-3}\,.
\label{nc}
\ee
In other words, the transition from superfluid core to isothermal normal region occurs when the density equals $n_{\rm c}$. 

As we move to larger radii in the normal region, the number density eventually becomes too low to sustain thermal equilibrium. At that point
we expect the profile to revert to a collisionless profile, such as the NFW profile. In a recent paper~\cite{Berezhiani:2017tth}, the matching
to the NFW profile was done explicitly by imposing suitable junction conditions. The analysis of~\cite{Berezhiani:2017tth} focused on the DM
superfluid effective theory relevant for MOND~\cite{Berezhiani:2015bqa,Berezhiani:2015pia}, and included a realistic treatment of the
baryon distribution.

In this work we ignore the NFW envelope and focus on the superfluid/normal region, where thermal equilibrium is justified.
Per~\cite{Berezhiani:2017tth}, this assumption is valid particularly for low-surface brightness galaxies,
where most of the mass is in the condensed phase. For concreteness, our prototypical galaxy has parameters 
\be
R_{\rm sf} = 10~{\rm kpc}\,;\qquad M_{\rm sf} = 4.7 \times 10^{9}~{\rm M}_\odot\,.
\label{MRSF}
\ee

\subsection{Density profile with 2-body interactions}

Let us start with the 2-body equation of state, derived in Sec.~\ref{2bodyeos}. To gain some ballpark intuition on the required
range of parameter values, it is instructive to study the density profile at $T=0$. Recall that the mean-field, zero-temperature equation of state is
given by the leading term in~\eqref{P02body}:
\be
P(T=0) \simeq \frac{g_2}{2}n^2 = \frac{2\pi\hbar^2a}{m} n^2\,,
\ee
where $a$ is the scattering length defined in~\eqref{scat length}. This approximation is valid for $a^3n\ll 1$. The hydrostatic equation~\eqref{hydrostatic_eqn} 
reduces to an $n=1$ Lane-Emden equation,
\be
\frac{1}{y^2} \frac{{\rm d}}{{\rm d}y}\left(y^2 \frac{{\rm d}\Xi}{{\rm d}y}\right) = -\Xi\,,
\ee
with dimensionless variables
\be
\Xi \equiv \frac{n}{n_0}\,;\qquad y \equiv \sqrt{\frac{G_{\rm N}m^3}{\hbar^2a}}\, r\,,
\ee
where $n_0$ is the central density. The exact solution, satisfying the boundary conditions $\Xi(0) = 1$ and $\Xi'(0) = 0$, is
\be
\Xi (y) = \frac{\sin y}{y}\,.
\label{density profile 2body T=0}
\ee

The profile terminates at $y = \pi$, which sets the core radius:
\be
R_{\rm sf} = \pi \sqrt{\frac{\hbar^2a}{m^3G_{\rm N}}} \simeq 18\, \sqrt{\frac{a}{\mu\text{m}}} \left(\frac{mc^2}{\rm 10\,eV}\right)^{-3/2}~{\rm kpc}\,.
\label{RSF 2body}
\ee
Meanwhile, the central density is related to the mass of the superfluid core via~\cite{chandrabook} $n_0 = \frac{\pi}{4} \frac{M_{\rm sf}}{R^3_{\rm sf}m}$, which gives
\be
n_0 \simeq \frac{M_{\rm sf}}{10^{9}\,{\rm M}_\odot} \left(\frac{a}{{\mu\text{m}}}\right)^{-3/2} \left(\frac{mc^2}{\text{10\,eV}}\right)^{7/2}\; 5.1 \times 10^5~{\rm cm}^{-3}\,.
\label{n02body_T=0}
\ee

We can now generalize the analysis to the finite-temperature case, including fluctuations. Substituting~\eqref{RSF 2body} for $R_{\rm sf}$ into~\eqref{haloT}, we find a DM temperature of
\be
T = \frac{M_{\rm sf}}{10^{9}~{\rm M}_\odot}~ \left(\frac{mc^2}{{\rm 10\,eV}}\right)^{5/2}\left(\frac{a}{{\mu\text{m}}}\right)^{-1/2}2 \times 10^{-5}~{\rm K} \,.
\label{T 2body}
\ee
The critical density $n_{\rm c}$ at this temperature is obtained by substituting for $R_{\rm sf}$ in~\eqref{nc}:
\be
n_{\rm c} =  \, \left(\frac{M_{\rm sf}}{10^{9}~{\rm M}_\odot}\right)^{3/2} \left(\frac{mc^2}{{\rm 10\,eV}}\right)^{21/4} \left(\frac{a}{{\mu\text{m}}}\right)^{-3/4}  \, 1.1 \times 10^3\,{\rm cm}^{-3}\,.
\label{nc 2body}
\ee

\begin{figure}[H]
     	\centering
        \includegraphics[height=3.in]{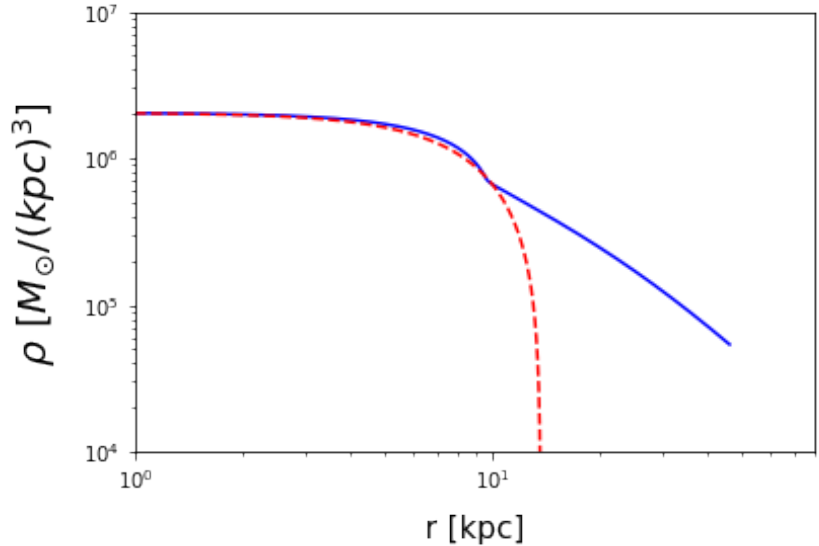}
    \caption{Density profiles for our finite temperature calculation (solid blue) and for the mean-field, $T=0$ calculation (dashed red). The phase transition occurs at $R_{\rm sf} = 10\ \text{kpc}$,
    where we notice a drop in density in the finite-temperature case.}
\label{density_2pt}
\end{figure}

To solve for the density profile, we evaluate the finite-temperature equation of state~\eqref{condensed phase pressure} at the temperature~\eqref{T 2body}, and substitute the result into 
the hydrostatic equation~\eqref{hydrostatic_eqn}. For concreteness, we choose the fiducial parameter values 
\be
a = 4.0\, \mu\text{m}\,;\qquad mc^2 = 25~{\rm eV}\,.
\label{m a fid}
\ee
Figure~\ref{density_2pt} shows the solution for the finite-temperature density profile (solid curve) for a superfluid region of mass $M_{\rm sf} = 4.7 \times10^{9}~{\rm M}_\odot$.
The temperature of the halo, according to~\eqref{T 2body}, is in this case $T = 3.25~{\rm mK}$. In practice, we substitute this temperature into the equation of
state~\eqref{condensed phase pressure}, and then adjust $n_0$ until the resulting profile matches the assumed mass $M_{\rm sf}$. This procedure gives
\be
n_0 = 1.57 \times 10^6\, \text{cm}^{-3}
\label{n03body_T=0}
\ee
Note that this central density together with the scattering length~\eqref{m a fid} satisfy the dilute Bose gas condition $a^3n\ll 1$.
From the plot, we notice a small drop in density at $R_{\rm sf} \simeq  10~{\rm kpc}$, which is where the phase transition takes place. For comparison, the dashed
curve shows the mean-field, $T=0$ density profile~\eqref{density profile 2body T=0}, for the same central density. 

Note that the fiducial parameters~\eqref{m a fid} blatantly violate the constraint on the scattering cross section per unit mass, $\frac{\sigma_{\rm scat}}{m} \;\gsim\; 0.5\;\frac{{\rm cm}^2}{{\rm g}}$, derived
from merging clusters~\cite{Markevitch:2003at,Randall:2007ph,Harvey:2015hha}. In terms of the scattering length, $\sigma_{\rm scat} = 4\pi a^2$, this bound
translates to
\be
a \;\lsim\; \sqrt{\frac{mc^2}{{\rm eV}}}~6 \times 10^{-5}\,{\rm fm}\,. 
\ee
It is easy to choose other values of $a$ and $m$ that satisfy the merging cluster constraint. For instance, $a = 3 \times 10^{-7}\,{\rm fm}$ and $mc^2 = 0.1~{\rm meV}$ satisfy the constraint and imply $R_{\rm sf} \simeq 10~{\rm kpc}$ as desired. It is worth keeping in mind, however, that this merging cluster constraint was derived in the context of self-interacting dark matter (SIDM)~\cite{Spergel:1999mh} and should be carefully revisited with superfluid DM.

\subsection{Density profile with 3-body interactions}

Moving on to the 3-body case, let us once again begin with the mean-field, zero-temperature profile, as originally derived in~\cite{Berezhiani:2015pia,Berezhiani:2015bqa}.
The equation of state in this approximation is given by the leading term~\eqref{P03body}:
\be
P(T=0) \simeq \frac{2}{3}g_3n^3 = \frac{\hbar^6}{12\Lambda^2m^3} n^3 \,.
\ee
where in the last step we have used~\eqref{g3 Lambda}. The hydrostatic equation reduces this
time to an $n=1/2$ Lane-Emden equation~\cite{Berezhiani:2015pia,Berezhiani:2015bqa},
\be
\frac{1}{y^2} \frac{{\rm d}}{{\rm d}y}\left(y^2 \frac{{\rm d}\Xi}{{\rm d}y}\right) = -\Xi^{1/2}\,,
\ee
where the dimensionless variables are now
\be
\Xi = \frac{n^2}{n_0^2}\,;\qquad y \equiv \sqrt{\frac{32\pi G_{\rm N} \Lambda^2m^5}{\hbar^6n_0}}\, r\,.
\ee
The boundary conditions are once again $\Xi(0) = 1$ and $\Xi'(0) = 0$.

The numerical solution is plotted in Fig.~4 of~\cite{Berezhiani:2015bqa}. A simple analytical form that
closely approximates the exact solution is~\cite{Berezhiani:2015pia,Berezhiani:2015bqa} 
\be
\Xi (y) \simeq \cos\left(\frac{\pi}{2}\frac{y}{y_1}\right)\,;\qquad y_1 \simeq 2.75\,.
\ee
In other words, 
\be
n(r) \simeq n_0 \cos^{1/2}\left(\frac{\pi}{2}\frac{r}{R_{\rm sf}}\right)\,.
\label{density profile 3body T=0}
\ee
The profile terminates at $y = y_1$, which sets the core radius:
\be
R_{\rm sf} = \sqrt{\frac{\hbar^6n_0}{32\pi G_{\rm N} \Lambda^2m^5}}\, y_1\,.
\label{RSF 3body}
\ee
The central density is related to the mass of the condensate region as follows~\cite{chandrabook}
\be
n_0 =  \frac{M_{\rm sf}}{4\pi R^3_{\rm sf}m} \frac{y_1}{|\Xi'(y_1)|} \,,
\label{n0 3body}
\ee
where, from the numerical solution, $\Xi'(y_1) \simeq -0.5$. 

Equations~\eqref{RSF 3body} and~\eqref{n0 3body} can be solved for $R_{\rm sf}$ and $n_0$ as a function of $M_{\rm sf}$:
\bea
\nonumber
R_{\rm sf} &\simeq& \left(\frac{M_{\rm sf}}{10^{9}\,{\rm M}_\odot}\right)^{1/5} \left(\frac{\Lambda\,m^3c^6}{10^{-3}~{\rm meV}\,({\rm 10\, eV})^3}\right)^{-2/5} \; 11~{\rm kpc}\,;\\
n_0 &\simeq& \left(\frac{M_{\rm sf}}{10^{9}\,{\rm M}_\odot}\right)^{2/5} \left(\frac{mc^2}{{\rm 10\,eV}}\right)^{-1} \left(\frac{\Lambda\,m^3c^6}{10^{-3}~{\rm meV}\,({\rm 10\, eV})^3}\right)^{6/5}  \; 4.0\times 10^{6}~{\rm cm}^{-3}\,.
\label{params 3body}
\eea
Note that $R_{\rm sf}$ depends on $M_{\rm sf}$ in this case, in contrast with the 2-body result~\eqref{RSF 2body}. 

\begin{figure}[H]
     	\centering
        \includegraphics[height=3.5in]{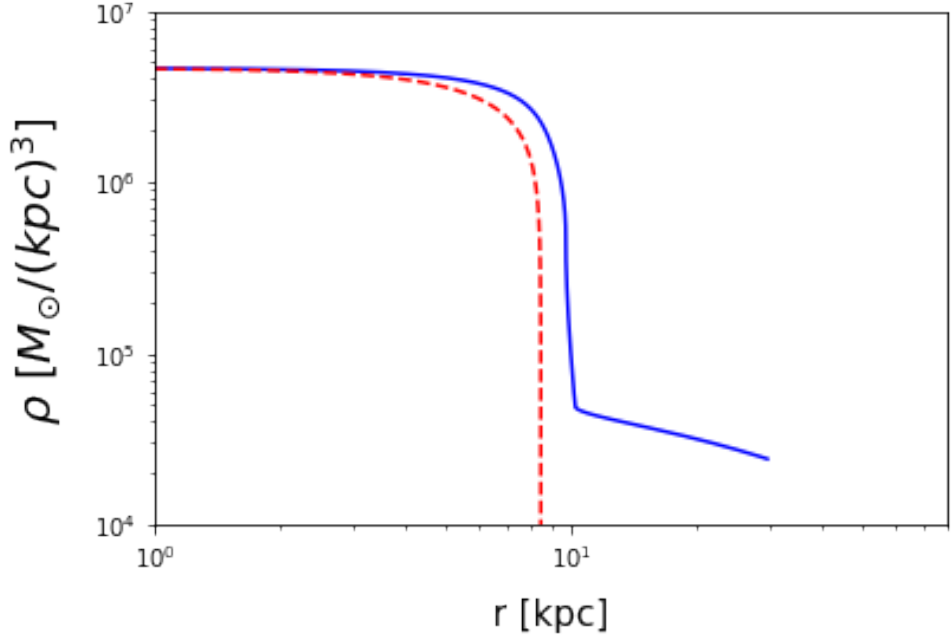}
    \caption{Density profiles for our finite temperature calculation (solid blue) and for $T = 0$ (dashed red). The phase transition occurs at $R_{\rm sf} = 10\ \text{kpc} $,
    where we notice a drop in density in the finite-temperature case.}
\label{density_3pt}
\end{figure}

We now generalize the analysis to the finite-temperature case, including fluctuations. Substituting the expression~\eqref{params 3body} for $R_{\rm sf}$ into~\eqref{haloT}, we find a DM temperature of
\be
T = ~\frac{mc^2}{{\rm 10\,eV}}~\left(\frac{M_{\rm sf}}{10^{9}~{\rm M}_\odot}\right)^{4/5} \left(\frac{\Lambda\,m^3c^6}{10^{-4}~{\rm meV}\,({\rm 10\, eV})^3}\right)^{2/5} 2.5\times10^{-3}\,~{\rm K} \,.
\label{T 3body}
\ee
The critical density $n_{\rm c}$ at this temperature is obtained by substituting for $R_{\rm sf}$ in~\eqref{nc}:
\be
n_{\rm c} =  \left(\frac{M_{\rm sf}}{10^{9}~{\rm M}_\odot}\right)^{6/5} \left(\frac{mc^2}{{\rm 10\,eV}}\right)^3  \left(\frac{\Lambda\,m^3c^6}{10^{-3}~{\rm meV}\,({\rm 10\, eV})^3}\right)^{3/5} 2.4 \times 10^{3}\,{\rm cm}^{-3}\,.
\label{nc 3body}
\ee
Using the temperature~\eqref{T 3body}, we evaluate the finite-temperature equation of state~\eqref{condensed phase pressure 3body} and substitute the result into the hydrostatic equation~\eqref{hydrostatic_eqn}. For concreteness, we choose the fiducial parameter values 
\be
mc^2 = 17~{\rm eV}\,;\qquad \Lambda = 1.2 \times 10^{-3}~{\rm meV}\,,
\label{m Lambda fid}
\ee

Figure~\ref{density_3pt} shows the solution for the finite-temperature density profile (solid curve) for $M_{\rm sf} = 1.0 \times 10^{10}~{\rm M}_\odot$, corresponds
to a halo temperature~\eqref{T 3body} of $T = 1.2~{\rm mK}$. As before, we adjust $n_0$ until the resulting profile matches the assumed mass $M_{\rm sf}$, with the result
\be
n_0 = 2 \times 10^4~\text{cm}^{-3} \,.
\label{n03body_T=0}
\ee
We notice a small drop in density at $R_{\rm sf} = 10\text{ kpc} $, which is where the phase transition takes place. For comparison, the dashed
curve shows the mean-field, $T=0$ density profile given by~\eqref{density profile 3body T=0}, for the same central density. 
It remains to check that the dilute Bose gas condition~\eqref{dilute 3body} is valid. A straightforward calculation gives
$\frac{1}{\pi}\left(\frac{mg_3}{\hbar^2}\right)^{3/4} n_0 \simeq 0.1$ for the fiducial parameters~\eqref{m Lambda fid}.

\section{Conclusions}
\label{conclusions}

In this paper we calculated the finite-temperature DM density profile, for superfluids with 2-body and 3-body interactions, using a self-consistent
mean-field approximation. The 3-body case serves as a toy model for the superfluid theory of~\cite{Berezhiani:2015pia,Berezhiani:2015bqa,Berezhiani:2017tth} 
to realize the MOND phenomenon. To simultaneously satisfy the mean-field self-consistency condition and ensure that the spectrum is gapless for $T \leq T_{\rm c}$,
we followed the  Yukalov--Yukalova approach based on two chemical potentials: one for the condensed phase, and another one for the normal phase.
Our calculation includes the contribution from the anomalous average, which is critical in overcoming the Hohenberg-Martin dilemma.

With knowledge of the equation of state $P = P(n,T)$, we derived the superfluid DM density profile in both 2-body and 3-body cases, for a fiducial galaxy with $M \sim 10^9~{\rm M}_\odot$. 
For simplicity we focused on static, spherically-symmetric halos and ignored the contribution of baryons. We also assume a constant temperature, $T = {\rm const.}$, consistent with thermal equilibrium. The resulting density profile  consists of a nearly homogeneous core, where the superfluid component dominates, surrounded by an isothermal ``atmosphere", where the normal component
dominates. The phase transition from superfluid core to isothermal normal region occurs when the density drops to the critical value.

Our results form the basis of a more detailed investigation of DM density profiles and explicit fits of galactic rotation curves, along the lines of~\cite{Berezhiani:2017tth}. 
In future work, we should include the contribution from realistic baryon distributions. Furthermore, we should consider the transition to the collisionless profile (such as the NFW profile),
which occurs when the density is too low to sustain thermal equilibrium. 

It would also be interesting to repeat the analysis for the more realistic superfluid effective theory proposed in~\cite{Berezhiani:2015pia,Berezhiani:2015bqa}.
As mentioned in Sec.~\ref{3body}, the hexic Hamiltonian~\eqref{H3body initial} gives a phonon action with the same power-law as the Bekenstein-Milgrom action,
but is off by a sign. A more complicated Hamiltonian was proposed in~\cite{Berezhiani:2015pia,Berezhiani:2015bqa} that gives the correct phonon action and
has stable perturbations. We plan to derive the self-consistent finite-temperature equation of state for that more realistic model in future work.

\vspace{.4cm}
\noindent
{\bf Acknowledgements:} We thank Lasha Berezhiani for helpful discussions, and Vivian Miranda for collaboration in the early stages. This work is supported by a New Initiative
Grant from the Charles E. Kaufman Foundation of the Pittsburgh Foundation.

\pagebreak
\renewcommand{\em}{}
\bibliographystyle{utphys}
\addcontentsline{toc}{section}{References}
\bibliography{Anu_paper_v10}

\end{document}